\documentclass{amsart}

\usepackage{amsmath,amsfonts}
\usepackage{algorithmic, color}
\usepackage{algorithm}
\usepackage{mathrsfs}
\usepackage{array}
\usepackage{subcaption}
\usepackage{textcomp}
\usepackage{stfloats}
\usepackage{url}
\usepackage{verbatim}
\usepackage{graphicx}
\newcommand{\qp}[1]{\ensuremath{\!\left({#1}\right)}}
\newcommand{\qb}[1]{\ensuremath{\!\left[{#1}\right]}}
\newcommand{\cS}{\ensuremath{\mathscr S}}
\newcommand{\up}{\Omega}
\newcommand{\e}{{\mathrm \epsilon}}
\newcommand{\dd}{{\mathrm d}}
\usepackage{xcolor}

\definecolor{amethyst}{rgb}{0.6, 0.4, 0.8}
\definecolor{applegreen}{rgb}{0.55, 0.71, 0.0}
\definecolor{aqua}{rgb}{0.0, 1.0, 1.0}
\definecolor{asparagus}{rgb}{0.53, 0.66, 0.42}
\definecolor{armygreen}{rgb}{0.29, 0.33, 0.13}
\definecolor{shitbrown}{rgb}{0.43, 0.21, 0.1}
\definecolor{brightpink}{rgb}{1.0, 0.0, 0.5}
\definecolor{brightube}{rgb}{0.82, 0.62, 0.91}
\definecolor{byzantine}{rgb}{0.74, 0.2, 0.64}

\usepackage{wrapfig}
\definecolor{fig}{RGB}{76,78,92}
\usepackage[labelfont=bf,font={color=fig,small}]{caption}

\usepackage{tikz}
\usetikzlibrary{intersections}
\usetikzlibrary{arrows,snakes,shapes}
\usepackage{pgfplots}
\usepgfplotslibrary{fillbetween}
                
\newcommand{\proton}[1]{%
    \shade[ball color=red] (#1) circle (.25);\draw (#1) node{$+$};
}
\newcommand{\neutron}[1]{%
    \shade[ball color=green] (#1) circle (.25);
}
\newcommand{\electron}[3]{%
    \draw[rotate = #3](0,0) ellipse (#1 and #2)[color=gray];
    \shade[ball color=yellow] (0,#2)[rotate=#3] circle (.1);
}
\newcommand{\nucleus}{%
    \neutron{0.1,0.3}
    \proton{0,0}
    \neutron{0.3,0.2}
    \proton{-0.2,0.1}
    \neutron{-0.1,0.3}
    \proton{0.2,-0.15}
    \neutron{-0.05,-0.12}
    \proton{0.17,0.21}
}

\newcommand{\inelastic}[2]{
  \proton{#1,#2};
  \draw[->,thick,blue](#1+0.5,#2)--(4,-3.7);%
  \draw[->,thick,blue](0,-3)--(-0.3,-4);%
  \shade[ball color=yellow] (-0.3,-4) circle (.1); 
}

\newcommand{\elastic}[2]{
  \proton{#1,#2};
  \draw[->,thick,orange,bend right=90](#1+0.5,#2) to  [out=-30, in=-150] (4,3.);
}

\newcommand{\protoncollision}[3]{
  \proton{#1,#2};
  \draw[->,thick,red](#1+0.5,#2)--(-0.5,0);%
  \draw[snake=coil, line after snake=0pt, segment aspect=0,%
    segment length=5pt,color=red!50!blue] (0,0)-- +(4,2)%
  node[fill=white!70!yellow,draw=red!80!white, below=.01cm,pos=1.]%
  {$\gamma$};%
  \draw[->,thick,red](#1+0.5,#2)--(-0.5,0);%
  \draw[->,thick,red](0.5,0)--(3.7,-1.8);%
  \neutron{4,-2};  
}
\usepackage{xspace}
\newcommand{\pg}{prompt-$\gamma$\xspace}
\renewcommand{\vec}[1]{\ensuremath{\boldsymbol{#1}}}

\usepackage{tikz, pgfplots}
\pgfplotsset{compat=newest}

\newcommand{\href}[1]{\url{#1}}

\begin{document}

\title[Bayesian Proton Therapy Verification]{A Bayesian Inverse Approach to Proton Therapy Dose Delivery
  Verification}

\author[Cox, Hattam, Kyprianou, Pryer]{
  Alexander M.~G.~Cox$^1$
  ,
  Laura Hattam$^2$
  ,
  Andreas E. Kyprianou$^3$
  ,
  Tristan Pryer$^{1,2}$.
}

\address{%
  $^1$Department of Mathematical Sciences, University of Bath, BA2 7AY
  \\
  $^2$Institute of Mathematical Innovation, University of Bath, BA2 7AY  
  \\
  $^3$Department of Statistics, University of Warwick, Coventry, CV4 7AL
}


\begin{abstract}
  This study presents a proof-of-concept for a novel Bayesian
  inverse method in a one-dimensional setting, aimed at proton beam
  therapy treatment verification. Our methodology is predicated on a
  hypothetical scenario wherein strategically positioned sensors
  detect \pg's emitted from a proton beam when it interacts with
  defined layers of tissue. Using this data, we employ a Bayesian
  framework to estimate the proton beam's energy deposition
  profile. We validate our Bayesian inverse estimations against a
  closed-form approximation of the Bragg Peak in a uniform medium and
  a layered lung tumour.
\end{abstract}

\maketitle

\section{Introduction}
Proton Beam Therapy (PBT) is a promising treatment for certain
challenging cancers. This includes cases where conventional
radiotherapy cannot adequately limit irradiation to surrounding
critical tissues. Examples are pediatric cancers, cancers at the base
of the skull, and complex head and neck cancers.

\begin{figure}[h!]
  \begin{center}
    \begin{tikzpicture}
    \begin{axis}[
        axis lines=left,
        domain=0:20,
        xmax=20,
        ymax=150,
        samples=100,
        xlabel=Depth (cm),
        ylabel=Relative dose,
        legend pos=north east,
        legend style={font=\tiny}, 
        grid=major,
        grid style={dashed,gray!30},
        ytick={0, 50, 100, 150}, 
        yticklabels={0, 50, 100, 150}, 
        y filter/.code={\pgfmathparse{#1*100}\pgfmathresult} 
    ]

    \addplot[color=blue, thick, smooth, mark=none] table [x index=0, y index=1, col sep=comma] {photon.csv};
    \addlegendentry{Photons (X-rays)}

    \addplot[color=red, thick, smooth, mark=none] table [x index=0, y index=1, col sep=comma] {carbon.csv};
    \addlegendentry{Carbon Ions}

    
    \addplot[color=green, thick, mark=none] table [x index=0, y index=1, col sep=comma] {proton.csv};
    \addlegendentry{Protons}

    \addplot[color=orange, thick, mark=none] table [x index=0, y index=1, col sep=comma] {SOBP_data.csv};
    \addlegendentry{Spread out Bragg Peak}

    \draw [dotted, thick] (axis cs:10,0) -- (axis cs:10,150);

    \draw [dotted, thick] (axis cs:15,0) -- (axis cs:15,150);

    \end{axis}
\end{tikzpicture}
  \end{center}
  \caption{\label{fig:sobp} An illustration of dose profiles of
    different radiation modalities. Notice that the dose deposition of
    photons is highest near the source with a decaying profile as
    depth increases. Protons and carbon ions are very localised with
    the decay of protons dropping to zero quickly after peak
    deposition is attained. This enables the construction of a Spread
    Out Bragg Peak over a target region by forming appropriate linear
    combinations of different energy and intensity beams. Carbon ions
    have an even sharper peak and dose profile upon entry but the
    deposition does not decay to zero after the peak due to secondary
    interactions.}
\end{figure}
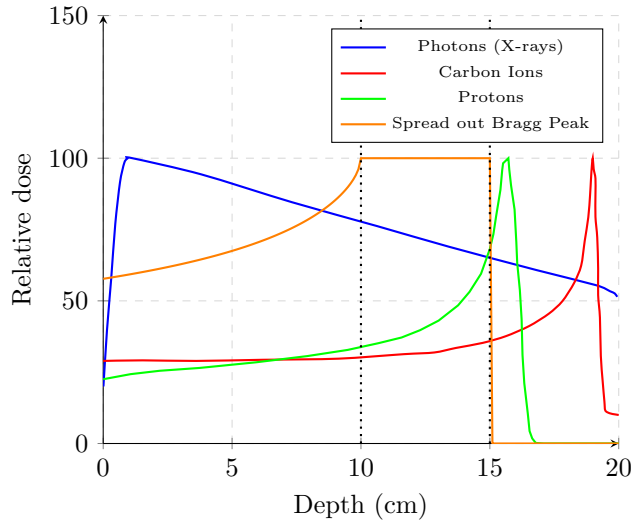

A proton deposits energy as it travels through matter. This energy
deposition increases as the proton slows down, reaching its maximum at
the end of its path, as illustrated in Figure \ref{fig:sobp}.

This energy, or dose, delivers the cancer-killing effect, but can also
harm surrounding healthy tissue. The objective of PBT is to administer
the intended dose to the tumor, as predetermined by prior clinical
evidence, while minimising exposure to the adjacent tissues. PBT's
potential for delivering a superior dose profile compared to
traditional photon treatments has long been recognised
\cite{lomax_intensity_2003}. A significant number of patients have
received PBT treatment
(\cite{particle_therapy_co-operative_group_ptcog_particle_nodate}
estimates over 275,000 worldwide as of December 2021). Nevertheless,
there remains a lack of conclusive evidence regarding PBT's benefits
over conventional photon-based treatments, such as X-rays (see, for
instance, \cite{chen_proton_2023}). As a relatively new treatment
modality, fundamental challenges persist in PBT's treatment planning
and verification, limiting its broader adoption. Day-to-day variations
in patient anatomy, such as those caused by water retention, introduce
uncertainties. Current strategies address these uncertainties during
planning either by expanding the irradiated volume - which diminishes
PBT's potential for sparing healthy tissue - or by compromising tumour
coverage to ensure protection of critical surrounding tissues.

In conventional photon radiotherapy, the dose delivery location can be
inferred by measuring the X-rays that exit the patient. However, while
proton therapy offers superior dose delivery, verifying its precise
administration is more challenging since most of the radiation dose,
including the proton beam itself, remains inside the patient. This
challenge, stemming from clinical need, is the focus of our
investigation, and in this work we propose a flexible mathematical
methodology which can form the basis for addressing this.

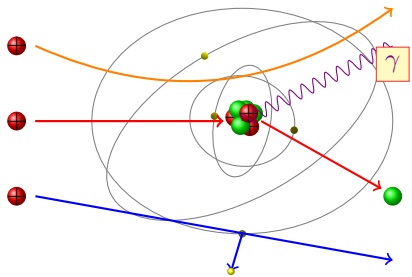
\begin{figure}[h!]
  \begin{center}
    \begin{tikzpicture}[scale=0.5]
      \nucleus \electron{1.5}{0.75}{80} \electron{1.2}{1.4}{260}
      \electron{4}{2}{30} \electron{4}{3}{180}
      \protoncollision{-6.}{0.}{160} \inelastic{-6.}{-2.}
      \elastic{-6.}{2.}
    \end{tikzpicture}
    \vspace{-5pt}
    \caption{\em The three main interactions of a proton with matter. A
      \textcolor{red}{nonelastic} proton-nucleus collision, an
      \textcolor{blue}{inelastic} Coulomb interaction with atomic
      electrons and \textcolor{orange}{elastic} Coulomb scattering with
      the nucleus.
      \label{fig:atom}
    }
  \end{center}
\end{figure}

When protons impact tissue, the primary mechanism of energy loss
involves inelastic interactions with atomic electrons. As the proton
interacts with electrons, it leaves behind a trail of excitations and
ionisations. There is a continual energy decrement of the proton
within the tissue, which is quantified by the stopping power. Elastic
interactions cause the treatment field to scatter within the tissue;
interestingly, this scattering increases with depth. Non-elastic
nuclear collisions can generate neutrons and other protons as
secondary interactions. An additional possible result of a nuclear
reaction is the decay of an unstable nucleus, which emits gamma rays,
or prompt gammas (\pg's). These are termed 'prompt' because they
emerge shortly after the initial proton interaction, as shown in
Figure \ref{fig:atom}.

The \pg's have energies in the range of approximately 2-10MeV. This is
significantly higher than the photon energies typically employed in
diagnostic imaging. As a result, there are currently no cameras
specifically optimised for measuring \pg's in a clinical setting.

The aim and novelty of this work lie in introducing a Bayesian
approach to explore the potential for dose delivery verification by
measuring the \pg radiation emitted from the patient. The location of
these emitted particles depends on the intricate interplay of the
environment they traverse, which often remains not fully understood
during treatment, given the limitations of tools like CT scans used
for planning. Consequently, determining whether the radiation measured
during the treatment aligns with the intended dose delivery poses a
complex inverse problem.

Our presentation revolves around the energy dynamics during proton
treatment. At the heart of our methodology is the Bortfield model
\cite{NewhauserZhang:2015,Bortfeld}, which we use to model forward
behaviour of a hypothetical scenario. We employ a Bayesian
framework to ascertain the potential success of treatments using
observed radiation profiles. In formulating the model we make some
fundamental assumptions. We suggest that the depth of energy
deposition correlates directly to the likelihood of a $\gamma$
particle's emission. Moreover, we consider our observational data to
be independent, identically distributed, and consistent over time.

A key aspect of our treatment is the data integration step. We adopt a
Bayesian approach, incorporating potentially highly uncertain beliefs
about the configuration of the patient, and synthesising this
information though our forward modelling to determine likely
configurations given the observed \pg{} emissions. Our approach takes
the form of a Bayesian Inverse Problem, which we propose to solve
using a Sequential Monte Carlo approach. This approach is now well
established, see for example
\cite{stuart_inverse_2010,beskos_sequential_2015,dunlop_bayesian_2016}. Benefits
of this approach are that we can control the computational burden of
solving the forward problem in a flexible framework which can handle
the statistical nature of the observed data.

Our goal is for the techniques proposed in this work to drive
advancements in next-generation measurement devices for PBT. The
potential of our approach extends beyond providing a sophisticated
solution to the intricate inverse problem of dose delivery
verification. It also quantifies the amount of information necessary
to yield accurate and reliable outcomes. In essence, this novel method
could shed light on the precision required in data gathering and
interpretation for effective dose delivery verification. This
understanding could, in turn, catalyse the development of advanced
imaging and detection systems, tailored specifically for the energies
and intricacies inherent to proton therapy. If realised, these
state-of-the-art devices could expand the current capabilities in PBT,
leading to enhanced treatments, better patient outcomes, and broader
adoption of this therapeutic technique.

The structure of the remainder of this paper is as follows: In \S
\ref{sec:FM}, we outline a prototypical forward model for the
study. \S \ref{sec:bayesian} delves into the Bayesian approach to
framing the inverse problem. In \S \ref{sec:KL}, the Kullback-Leibler
divergence is introduced, highlighting its application in
differentiating between different patient configurations. \S
\ref{sec:SMC} presents the Sequential Monte-Carlo methods and their
relevance to the Bayesian inverse problem. Lastly, \S
\ref{sec:numerics} showcases several numerical experiments.

\section{Forward Model}\label{sec:FM}

Our initial objective is to construct a robust forward model. In
essence, we aim to delineate how a given physical setup (termed
'patient configuration') influences proton transport, culminating in
the observed $\gamma$-radiation.

The computation of pristine Bragg curves, which chart the energy loss
of protons as they traverse matter, can be undertaken using a spectrum
of techniques. These range from lookup tables populated with
empirically measured data to Monte-Carlo simulations and analytical
models. In this section, we provide an overview of the foundational
physics-based models, beginning with the articulation of a general
partial differential equation governing the transport of charged
particles.

While the mathematical formulation capturing proton transport in its
full breadth is seldom explicitly stated, it has recently been
addressed comprehensively by Birmpakos et.\ al.\ \cite{BKP}. Broadly
speaking, one has to navigate the spatial-velocity-energy phase
space. To facilitate this, we define $X\subset \mathbb R^3$ as a
closed, bounded spatial domain, $\mathbb S^2$ as the unit sphere, and
$E = (\mathtt{e}\mathtt{min},
\mathtt{e}\mathtt{max})\subset(0,\infty)$. This paves the way to
define the spatial-velocity-energy phase space as
\begin{equation}
  \Upsilon = X\times \mathbb{S}^2 \times E.
\end{equation}

The mechanism by which we will account for sources stems from the
spatial boundary. Let us represent it as:
\begin{equation}
  \partial X_-
  :=
  \{x\in\partial X : \Omega \cdot {\vec n_X} \leq 0\}
\end{equation}
where ${\vec n_X}$ will serve as the outward-pointing normal to
$\partial X$. We postulate that protons are introduced into the system
through the inflow boundary $\partial X_-$ at an initial flux denoted by
$g$.

The transport equation that dictates proton flux, $\psi$, is
occasionally termed the Proton Transport Equation
(PTE). Alternatively, it's known as the Boltzmann--Fokker--Planck
Equation satisfying
\begin{align}
  \partial_t \psi
   +
   \overbrace{  \Omega \cdot \nabla_{D} \psi}^{\text{transport}}
   -
   \overbrace{\nabla_\varepsilon \qp{\varsigma \psi}}^{\text{energy depletion}} 
   -\overbrace{\mu \Delta_{  \Omega} \psi}^{\text{elastic scatter}}
   =
   \overbrace{\cS \psi}^\text{scattering}
  \label{eq:BTE}
   \end{align}
   in $(0,T] \times \Upsilon$,
   with boundary conditions
\begin{align}
 \psi(\vec x, 0) &= \psi_0(\vec x) \text{ in }\{0\} \times \Upsilon\notag\\
   \psi &= \underbrace{g}_{\text{proton source}} \text{ on }  (0,T] \times \partial X_-.
 \label{eq:BTEbcd}
 \end{align}
 Here, the scatter operator $\cS$, incorporating elastic and inelastic
 scatter, satisfies
\begin{align*}
  \cS \psi
  & :=
  \int_{S^2 \times E}
  \kappa(y, \up', \e')
  \pi(y, \up', \up, \e', \e)
  \psi(y, \up', \e') \dd \up' \dd \e' -
  \kappa(y, \up, \e) \psi(y, \up, \e),
\end{align*}
where $\kappa$ is the scatter cross section and $\pi$ is the
distribution of the new direction and energy on a scatter event. 

In this general framework, the energy released into the tissues is
determined by the energy release from non-elastic scatter events as
detailed in \eqref{eq:BTE}. However, to date, no such calculations
have been documented in the literature. This omission represents just
one of the many gaps in the broader mathematical framework, which has
yet to be fully developed to facilitate comprehensive dose
calculations.

Due to the constraints of brevity and to simplify exposition we will
explore the Bayesian framework with an analytical model. This model
encapsulates many crucial physical processes and offers a combination
of computational simplicity, speed, and practical utility for our
context. We are interested in a one dimensional depth-dose
representation over the spatial domain $x\in X$ where we suppose that
a series of adjacent $\gamma$-detectors positioned above the patient
at a distance $h$. We appeal to the work of Bortfeld \cite{Bortfeld}
where the author derives a closed form expression for a dose-depth
curve along a single pencil beam under a number of basic physical
assumptions. We can use this to calibrate a forward model for a
particular energy proton beam in depth, from which we can generate
synthetic data of $\gamma$ readings from our hypothetical
$\gamma$-detectors, sometimes known as a Compton camera.

Let $D(x)$ denote the dose and suppose $\Phi_0$ denotes
primary fluence, $R$ the range, $\sigma$ the standard deviation of the
Gaussian of depth, $\zeta(x) = \tfrac{R - x}{\sigma}$, $\epsilon$ is
the fraction of low-energy proton fluence to total fluence, $\Gamma$
is the standard $\Gamma$ function and $\mathcal D_y(x)$ is the
parabolic cylinder function. Then further, let $\alpha$, $p$ and
$\rho$ be material dependent constants from the Bragg-Kleeman rule for
stopping power
\begin{equation}
  \label{eq:Bragg-Kleeman}
  -\frac 1 \rho \frac{d E}{d x} \approx \frac{E^{1-p}}{\rho \alpha p}.
\end{equation}

The Bortfeld model then consists of writing the dose function $D$ as
\begin{equation}
  \label{eq:Bortfeld}
  D(x)
  =
  K_1
  \qb
     {
       \frac 1 \sigma \mathcal{D}_{-1/p}(-\zeta(x))
       +
       K_2
       \mathcal{D}_{-1/p-1}(-\zeta(x))
     }
\end{equation}
with
\begin{equation}
  \begin{split}
    K_1&:=\Phi_0
    \frac
        {\exp\qp{-\zeta(x)^2/4} \sigma^{1/p} \Gamma(1/p)}
        {\sqrt{2\pi}\rho p \alpha^{1/p}(1+\beta R)}
        \\
        K_2&:=
        \qp{
          \frac \beta p
          +
          \hat \gamma \beta
          +
          \frac \epsilon {R}
        }.
  \end{split}
\end{equation}
Although strictly 1-dimensional, this model exhibits remarkable
accuracy when juxtaposed with measured dose values and aligns with the
objectives of our study. Indeed, this expression has been employed in
treatment planning calculations \cite[c.f.]{SzymanowskiOelfke:2003}.

In our context, the inputs for the Bortfeld model comprise a set of
$n$ material-dependent constants. For convenience we highlight the
constants physical meaning and typical values in Table
\ref{tab:coeff}. We make the assumption that some of these constants,
specifically $R, \sigma, \epsilon$ are unknown in the tissue of the
patient. We denote $\vec{d} = \qp{R, \sigma, \epsilon}$ and where we
want to make the dependency of the dose ${D}$ on $\vec{d}$ explicit,
we will write ${D}(x | \vec{d})$. Furthermore, in later sections we
construct a stratified representation of a patient, as depicted in
Figure \ref{fig:lung} and represent write $\vec d = \{ R_i, \sigma_i,
\epsilon_i \}_{i=1}^n$. The values of these coefficients are
contingent upon the patient's geometry since they correlate with the
medium inherent to each region.

  \begin{table}[h!]
    \begin{center}
      \begin{tabular}{c|c|c}
      Constant & Meaning & Value \\ \hline R & Range & TBD \\ $\sigma$
      & Width of Gaussian & TBD \\ $\epsilon$ & Fraction of primary
      fluence contributing to the ``tail'' of the energy spectrum &
      TBD \\ $\alpha$ & Proportionality factor & 0.0022 \\ p &
      Exponent of Bragg Kleeman range-energy rule & 1.77 \\ $\rho$ &
      Density & 1 \\ $ \beta $& Slope of fluence reduction relation &
      0.012 \\ $ \hat{\gamma}$ & Fraction of locally absorbed energy
      released in nonelastic nuclear interactions & 0.6 \\ \hline
    \end{tabular}
    \end{center}
    \caption{\label{tab:coeff} Summary of constants and parameters
      used in the theoretical model. These are extracted from
      \cite{Bortfeld} for a uniform water phantom.}
  \end{table}

\begin{figure*}[h!]
  \centering
  \includegraphics[width=0.3\textwidth]{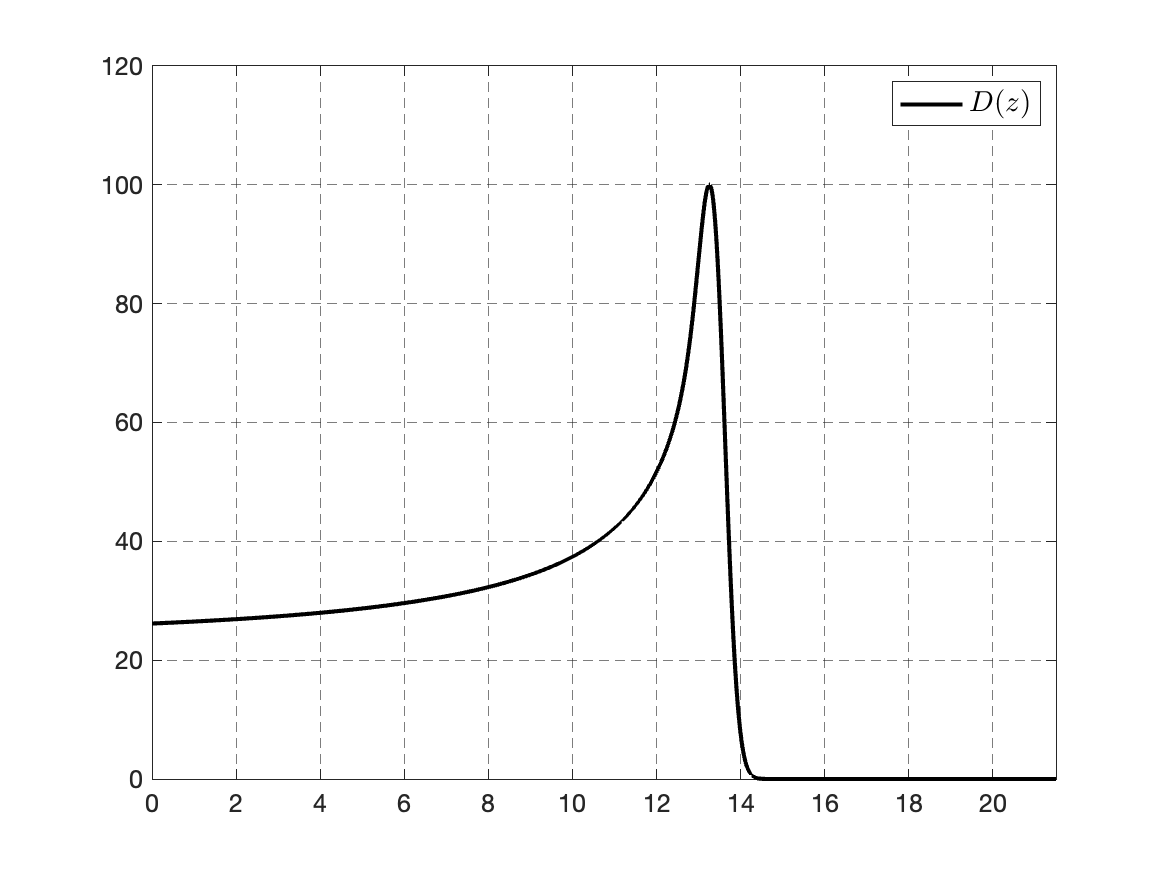}  
  \includegraphics[width=0.3\textwidth]{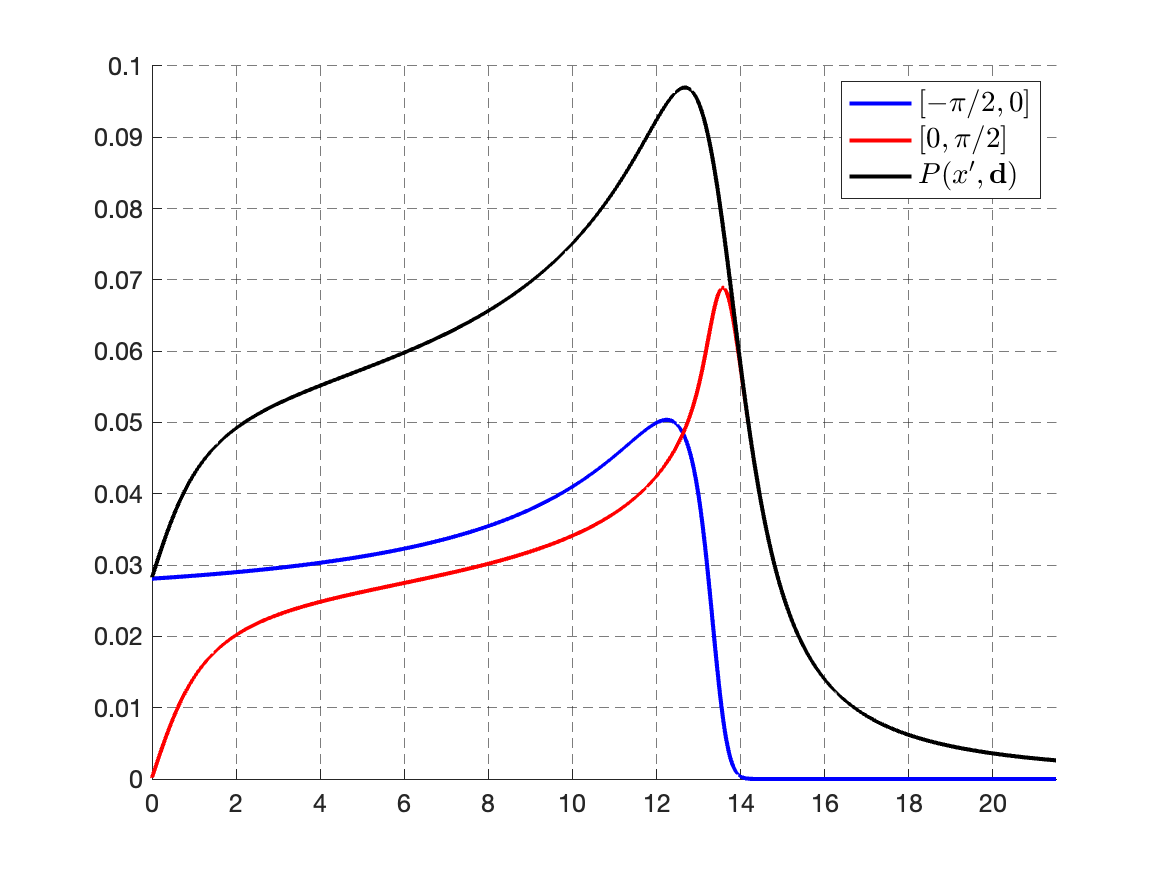}
  \includegraphics[width=0.3\textwidth]{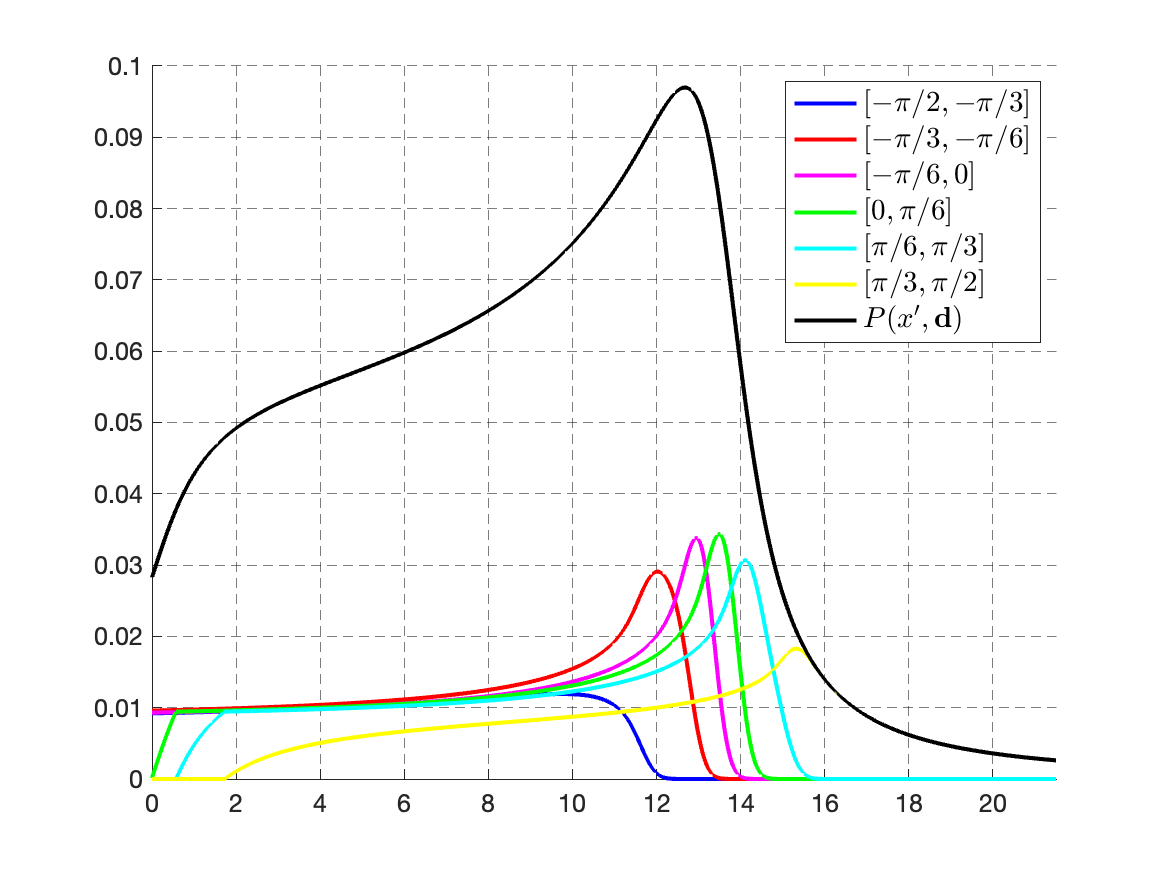}
  \caption{
  \label{binning}
  A Bragg Peak (left) and two examples of how the \pg distribution
  $P(x'|\vec{d})$ can be split into two (middle) and six (right)
  distributions of $\gamma$-detection distributions from coding
  different emission ranges. In both cases the detectors are
  positioned at distance $h=1$.}
\end{figure*}

We then posit the following assumption about the likelihood of an
isotropic emission of a \pg particle from a given location:

\begin{quote}
  {\bf Assumption:} The probability of a $\gamma$-particle being
  emitted from a point at depth $x\in X$ along the pencil beam is
  proportional to the energy deposition from the Bortfield model. 
\end{quote}
This assumption is visualised in Figure \ref{binning}.

An immediate consequence of this assumption is that we can specify the
law of $\gamma$-particles emitted in terms of the dose function $D$
given in \eqref{eq:Bortfeld}. Specifically, for a given set of
constants $\vec{d}$, the distribution of the emission point of
$\gamma$-particles is given by
\begin{equation}
  \label{eq:Qdefn}
  Q(x|\mathrm{d}) = \frac{\mathcal{D}(x|\mathrm{d})}{\int
    \mathcal{D}(y|\mathrm{d})\, dy}
\end{equation}

Using $Q$, the probability distribution that a $\gamma$-particle lands
at a certain location along the detectors can be determined. We first
consider the setup of the problem.

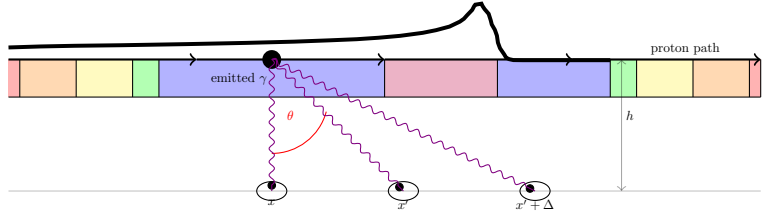
\begin{figure*}[h!]
  \centering
  \begin{tikzpicture}[scale=0.5]
    \def\h{7}  
    \def\posx{7}  

    \def\layerMin{1.5} 
    \def\layerHeight{2.5} 
    
    \fill[red!30] (0,\layerMin) rectangle (0.3,\layerHeight);
    \fill[orange!30] (0.3,\layerMin) rectangle (1.8,\layerHeight);
    \fill[yellow!30] (1.8,\layerMin) rectangle (3.3,\layerHeight);
    \fill[green!30] (3.3,\layerMin) rectangle (4.0,\layerHeight);
    \fill[blue!30] (4.0,\layerMin) rectangle (10.0,\layerHeight);
    \fill[purple!30] (10.0,\layerMin) rectangle (13.0,\layerHeight);
    \fill[blue!30] (13.0,\layerMin) rectangle (16.0,\layerHeight);
    \fill[green!30] (16.0,\layerMin) rectangle (16.7,\layerHeight);
    \fill[yellow!30] (16.7,\layerMin) rectangle (18.2,\layerHeight);
    \fill[orange!30] (18.2,\layerMin) rectangle (19.7,\layerHeight);
    \fill[red!30] (19.7,\layerMin) rectangle (20.0,\layerHeight);

    \foreach \depth in {0.3,1.8,3.3,4.0,10.0,13.0,16.0,16.7,18.2,19.7,20.0}{
      \draw (\depth,\layerMin) -- (\depth,\layerHeight);
    }
        
    \draw (0,\layerMin) -- (20,\layerMin);

    \draw[->, thick] (0,2.5) -- (5,2.5);
    \draw[->, thick] (5,2.5) -- (10,2.5);
    \draw[->, thick] (10,2.5) -- (15,2.5);
    \draw[->, thick] (15,2.5) -- (20,2.5);
    \node at (18,2.5) [above,scale=0.5] {proton path};
    
    \fill (\posx,2.5) circle (0.25cm);
    \node at (\posx,2) [left,scale=0.5] {emitted $\gamma$};

    \def\x{-1}  
    \def\posy{2.5} 
    \def\x{-1}  

    \def\startAngle{270} 
    \def\endAngle{345} 
    
    \draw[red] (\posx,1.5) ++(\startAngle:1.5) arc (\startAngle:\endAngle:1.5);
    \node[red, scale=0.5] at (\posx + 0.5, 1.) {$\theta$};
    
    \foreach \i/\j in {1/$x$,1.5/$x'$,2/{$x' + \Delta$}}{
        \def\eyex{\h*\i}
        \def\eyey{\x}
        
        \pgfmathsetmacro{\angle}{atan2(\posy-\eyey,\posx-\eyex)}
        \pgfmathsetmacro{\pupilx}{\eyex + 0.15*cos(\angle)} 
        \pgfmathsetmacro{\pupily}{\eyey + 0.15*sin(\angle)} 

        \draw[fill=white] (\eyex,\eyey) ellipse (0.4cm and 0.25cm); 
        \draw[fill=black] (\pupilx,\pupily) circle (0.1cm); 

        \draw[decorate, decoration={coil, aspect=0, segment length=5pt, amplitude=1pt}, color=red!50!blue] 
              (\posx,\posy) -- (\eyex,\eyey);
        
        \node at (\eyex,\eyey-0.1) [below,scale=0.5] {\j};
    }

    \begin{axis}[
        hide axis,
        scale only axis,
        width=16cm, 
        height=10cm, 
        xmin=0, xmax=20,
        ymin=-1.65, ymax=5, 
    ]
      \addplot [
        color=black, 
        line width=3pt, 
        mark=none        
        ] table [
            x index=0, 
            y index=1, 
            col sep=comma
        ] {proton.csv};    
    \end{axis}    

    \def\bottomAxis{-1.} 
    
    \draw[thin, opacity=0.25] (0,\bottomAxis) -- (20,\bottomAxis); 

    \draw[thin, opacity=0.35, <->] (16.3,\bottomAxis) -- (16.3,2.5); 
    \node at (16.3,1) [right,scale=0.5] {$h$};
    
\end{tikzpicture}
  \caption{Simulating $\gamma$-particles: $\gamma$-particles are released from the $x$-axis due to the deposited dose. Some of these particles land on the series of adjacent $\gamma$-detectors positioned along the $x'$-axis. These axes are separated by the distance $h$. \label{pc}}
\end{figure*}

The location of the deposited dose ($0\leq x\leq 1$) relative to the
series of $\gamma$-detectors ($0\leq x'\leq 1$) is illustrated in
Figure \ref{pc}. With the help of this figure, it can be shown that
the probability of a $\gamma$-particle landing at the position
$[x',x'+\Delta)$, given that it started at position $x$, at a uniform
angle in $[0,2 \pi]$, is defined as
\begin{equation}
  P(x'|x)
  =
  \frac{1}{2\pi}\qp{
    \tan^{-1}\qp{
      \frac{x'+\Delta-x}{h}
    }
    -
    \tan^{-1}\qp{
      \frac{x'-x}{h}
    }
  },
\end{equation}
where $\Delta$ is the `width' of the detector. Therefore, the
probability of the $\gamma$-particle landing in $[x',x'+\Delta)$, for
some choice of $\vec{d}$, is given by
  \begin{align}
    P(x'|\vec{d})
    &=
    \int P(x'|x)Q(x|\vec{d})dx \nonumber\\
    &=
    \int \frac{Q(x|\vec{d})}{2\pi}
    \times\left(\tan^{-1}\left(\frac{x'+\Delta-x}{h}\right)-\tan^{-1}\left(\frac{x'-x}{h}\right)\right)
      dx, \label{dprobx}
  \end{align}
where $Q(x|\vec{d})$ corresponds to the probability distribution
for the origin of the $\gamma$-particles, which is the dose profile
generated by the analytical model with input $\vec{d}$. 

Hence, for some given patient geometry that defines $\vec{d}$, the
analytical model is computed to obtain the dose profile,
$Q(x|\vec{d})$, which is used in
\eqref{dprobx}. As a result of the deposited dose, $\gamma$-particles
are released, with some landing on the $\gamma$-detectors. To simulate
these observations along the detectors, we randomly sample from
(\ref{dprobx}).

Our forward problem can then be defined as
\begin{equation}
  P(\cdot | \vec{d}) = F(\vec{d}),
\end{equation}
where $F$ represents the analytical model of the proton beam and
$\gamma$-particle emissions with $\vec{d}$ representing the model
inputs. The model output $P(\cdot| \vec{d})$ is the distribution of
the position of $\gamma$-particles detected. We assume in addition
that, given the model inputs $\vec{d}$, each particle detection
location is independent.

As an extension to the $\gamma$-particle simulation model, we suppose
that the detectors can determine a projection angle range for each
detected particle. The projection angle that is estimated by the
detectors is labelled $\theta$ in Figure \ref{pc}. The detectable
angle ranges are defined as
$[-\pi/2,-\pi/2+\delta\theta),[-\pi/2+\delta\theta,
-\pi/2+2\delta\theta),\ldots,[\pi/2-\delta\theta,\pi/2)$, where
$\delta \theta=\pi/b$ and the model parameter $b$ specifies the number
of bins. The distribution used to simulate $\gamma$-particle
observations, (\ref{dprobx}), is then split into $b$ bins according to
these angle ranges, where
$P(x'|\vec{d})=\sum_{i=1}^{b}P_i(x'|\vec{d})$ and $P_i$ are
the `binned' probability distributions. To now generate $\gamma$
readings along the detectors, we firstly sample
$i\in \{1,2,\ldots,b\}$ with probability $p_i = \int_0^1
P_i(x|\vec{d}) \, dx$ respectively, and next
randomly sample from bin $i$'s probability distribution
$P_i(x'|\vec{d})$. As an example, the partitioning of the
distribution into $2$ and $6$ separate bins is demonstrated in Figure
\ref{binning}.

\section{Bayesian Inverse Problem}
\label{sec:bayesian}

From an operator's perspective, determining whether the radiation
observed by the detectors aligns with a specific patient configuration
is crucial. We advocate for a Bayesian modeling approach. Typically,
the operator would suggest a prior model for the actual patient
configuration during therapy, potentially based on scans taken during
the treatment planning phase. After defining this prior, our aim is to
utilize the observed radiation profile to assess the likelihood of a
successful treatment. As detailed earlier, given a patient
configuration, the data distribution can be derived by solving the
forward problem. However, we anticipate that solving this forward
model with high precision will be numerically demanding. Consequently,
we need efficient numerical methods to update our potentially
high-dimensional and intricate posterior distribution based on data
observations.

We make an assumption that our observations are independent and
identically distributed, and the model remains static over time. This
challenge falls within the realm of sequential Bayesian learning
problems (e.g., \cite[Section~3.2]{chopin_introduction_2020}). Our
proposal leans towards a model using Sequential Monte Carlo (SMC)
methods. Considering the likelihood of configurations being complex
and high-dimensional - capturing potential deformations, uncertainties
in patient positioning, physical property variations, and more - we
aim to design Bayesian methods that explore the prior space
effectively while necessitating minimal solutions of the system for
specified configurations. In our SMC approach, the posterior measure
is represented by a 'cloud' of potential points, updated based on
incoming data. We label these elements in the 'cloud' as particles;
each particle represents a possible patient geometry configuration,
requiring a numerical solution of the forward problem.

Broadly, our algorithm, elaborated on further in Section~\ref{sec:SMC}
below, adheres to an importance-selection-mutation pattern. In the
importance phase, observed data is harnessed to update the posterior
probability linked to each particle: particles predicting the data
accurately will see a relative boost in their posterior probability,
while the others will see a decrease. The selection phase involves
rebalancing the particle population-eliminating unlikely particles and
duplicating promising ones. The mutation phase introduces particle
modifications based on stochastic dynamics. By comparing new and
existing particles, an accept-reject step is executed based on
prediction accuracy. Without mutation, particles would quickly
converge to a few models, hindering the assimilation of extensive data
for model refinement. Given the crucial role of mutation, it might be
practical to process data in smaller 'chunks', despite data collection
potentially occurring within a brief timeframe. Thus, we suggest a
block-sequential data analysis approach, facilitating efficient
interplay between selection and mutation.

This method yields a Bayesian posterior probability measure describing
the distribution of $\vec{d}$ given the observed data. Ideally,
this could offer insights on treatment outcomes-like posterior
estimates of treatment success, potential damage to high-risk zones,
or total energy deposition.

For the scope of this paper, we confine ourselves to simpler models
with a few input parameters delineated in the preceding section. Our
goal is to harness data from the $\gamma$-detectors to estimate values
of $\vec{d}$, subsequently predicting the medium's properties ---
possibly tumor position --- and dose distribution in the vicinity. By
framing this as a sequential Bayesian learning challenge, the
posterior probability of a particular configuration $\vec{d}$ can be
articulated when new data points are encountered.  Suppose that we
observe $k$ $\gamma$-particles at locations
$\vec{x}' = (x_1',\dots, x_k')$, and want to use this data to infer
the likely values of $\vec{d}$. We can proceed as follows. Using
Bayes' Theorem, the posterior distribution is given as:
\begin{equation}
  p(\vec{d}|\vec{x}')
  =
  \frac{p(\vec{x}'|\vec{d})p(\vec{d})}{p(\vec{x}')},\label{bt}
\end{equation}
where $p(\vec{x}'|\vec{d})$ is the likelihood function,
$p(\vec{d})$ is the prior distribution and
$p(\vec{x'})=\int_{\vec{d}}
p(\vec{d}|\vec{x'})p(\vec{d})d\vec{d}$ is the marginal
likelihood. In general, $p(\vec{x'})$ is not easily computed, and
SMC methods, as described in Section~\ref{sec:SMC} are in large part
motivated by the need to avoid the challenge of computing this
quantity.

For the prior distribution, we can apply a Gaussian white noise
prior on the parameters of $\vec{d} = \{R_i, \sigma_i, \epsilon_i\}_{i=1}^n$ 
where the mean and covariances could be chosen to
reflect some prior knowledge about $\vec{d}$ before any
observations are made.

The likelihood function is defined using (\ref{dprobx}) since this
equation gives the probability of observing the location $x_i'$, given
a particular choice of the model parameters $\vec{d}$. Moreover,
for $k$ new independent observations, the likelihood is given by
\begin{equation*}
p(\vec{x}'=(x_1',x_2',\ldots,x_k')|\vec{d})=\prod_{j=1}^{k}P(x_j'|\vec{d}).
\end{equation*}

Thus, we begin with limited information about the model parameters
$\vec{d}$, represented by the prior. Next, using (\ref{bt}), the
posterior is updated by adding $\gamma$-particle readings
$\vec{x}'$. This distribution reveals the most likely state of
$\vec{d}$ given $\vec{x}'$ which, at least usually, will become
more refined as the number of observations increases, and therefore,
giving us a well-informed estimate for $\vec{d}$. Once these model
parameters are established within a certain level of confidence, the
physical parameters and the corresponding dose profile can be
approximated.


\section{KL-based approach to model discrimination}
\label{sec:KL}

A question of paramount importance to address before practically
implementing our methods is the feasibility of distinguishing between
two possible patient configurations, especially considering that the
data collection will be restricted by the duration of the proton
treatment. Further, it's worth questioning whether enhancing the
measurements--by, for instance, increasing the number or sensitivity of
gamma detectors or integrating angular sensitivity-would augment our
capability to differentiate between successful and unsuccessful
treatments. To address this, we employ the Kullback-Leibler (KL)
divergence, an information-theoretic perspective on relative entropy,
with a focus on its interpretation as discrimination information
(\cite{kullback_information_1951}).

To simplify our understanding, let us consider wanting to discriminate
between just two potential physical configurations. The fundamental
query here is: how much data is required to accomplish this? By
contemplating two distinct geometries that we might wish to
differentiate, we can, in fact, provide estimates on the necessary
caliber of the detector setup.

Let us assume we have two closely related probability distributions,
$P(x'|\vec{d}_t)$ and $P(x'|\vec{d}^*)$, which are the
likelihood functions for the true solution and a neighbouring solution
respectively. In order to distinguish between the two distributions,
the number of observations, $k$, sampled from $P(x'|\vec{d}_t)$
must satisfy
\begin{align}
k&\ge \left(\log\left(\frac{1-p_0}{p_0}\right)-\log\left((1-\delta)^{-1}-1\right)\right)\nonumber\\
&\quad\quad\quad\quad\quad\quad\times D_{KL}(P(x'|\vec{d}_t)|P(x'|\vec{d}^*))^{-1},\label{klc}
\end{align}
where
\begin{align*}
  D_{KL}(P(x'|\vec{d}_t)|P(x'|\vec{d}^*))
  &=
  \int\log \left(\frac{P(x'|\vec{d}_t)}{P(x'|\vec{d}^*)}\right)
  \times P(x'|\vec{d}_t)dx'
\end{align*}
is the Kullback-Leibler (KL) divergence, $p_0$ is the prior
probability that $\vec{d}_t$ is true, and $0< \delta \ll 1$ is an
acceptable error rate. If the likelihood functions have been split
into $b$ bins such that
$P(x_j'|\vec{d})=\sum_{i=1}^{b}P_i(x_j'|\vec{d})$, the KL
divergence is instead written as
\begin{align*}
D_{KL}(P(x'|\vec{d}_t)|P(x'|\vec{d}^*)) =\sum_{j=1}^b\int
  \log\left(\frac{P_j(x'|\vec{d}_t)}{P_j(x'|\vec{d}^*)}\right) \times P_j(x'|\vec{d}_t)dx'.
\end{align*}

If (\ref{klc}) is met, then the posterior probability of the true
solution will satisfy
\begin{equation*}
  p(\vec{d}=\vec{d}_t|x_1',x_2',\ldots,x_k')\ge 1-\delta,
\end{equation*}
with probability of order 1, which means the truth can be estimated
with a high level of confidence.

Suppose we are considering two closely related patient geometries:
$\vec{d}_t$ and $\vec{d}^*$. If $\vec{d}_t$ represents the
true solution, then to distinguish between these two configurations,
the number of observations, $k$, must satisfy the condition given by
equation (\ref{klc}). This criterion is influenced by the number of
bins, $b$, that the $\gamma$-detectors can discern. Additionally, the
sensitivity of the forward model to variations in each of the
parameter values plays a crucial role.

\begin{figure*}[h!]
\centering
\begin{subfigure}{0.4\textwidth}
  \includegraphics[width=\textwidth]{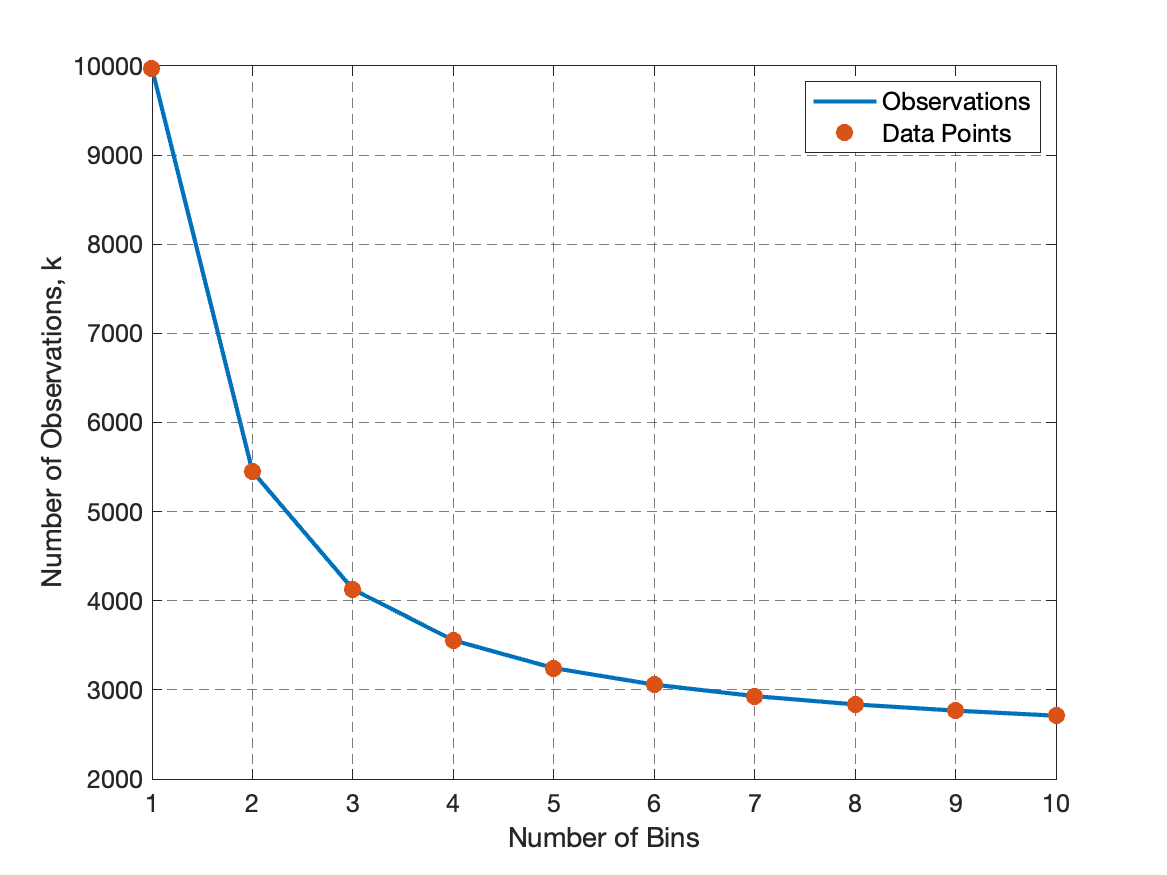}
  \caption{$h=0.5$.}
\end{subfigure}
\begin{subfigure}{0.4\textwidth}
  \includegraphics[width=\textwidth]{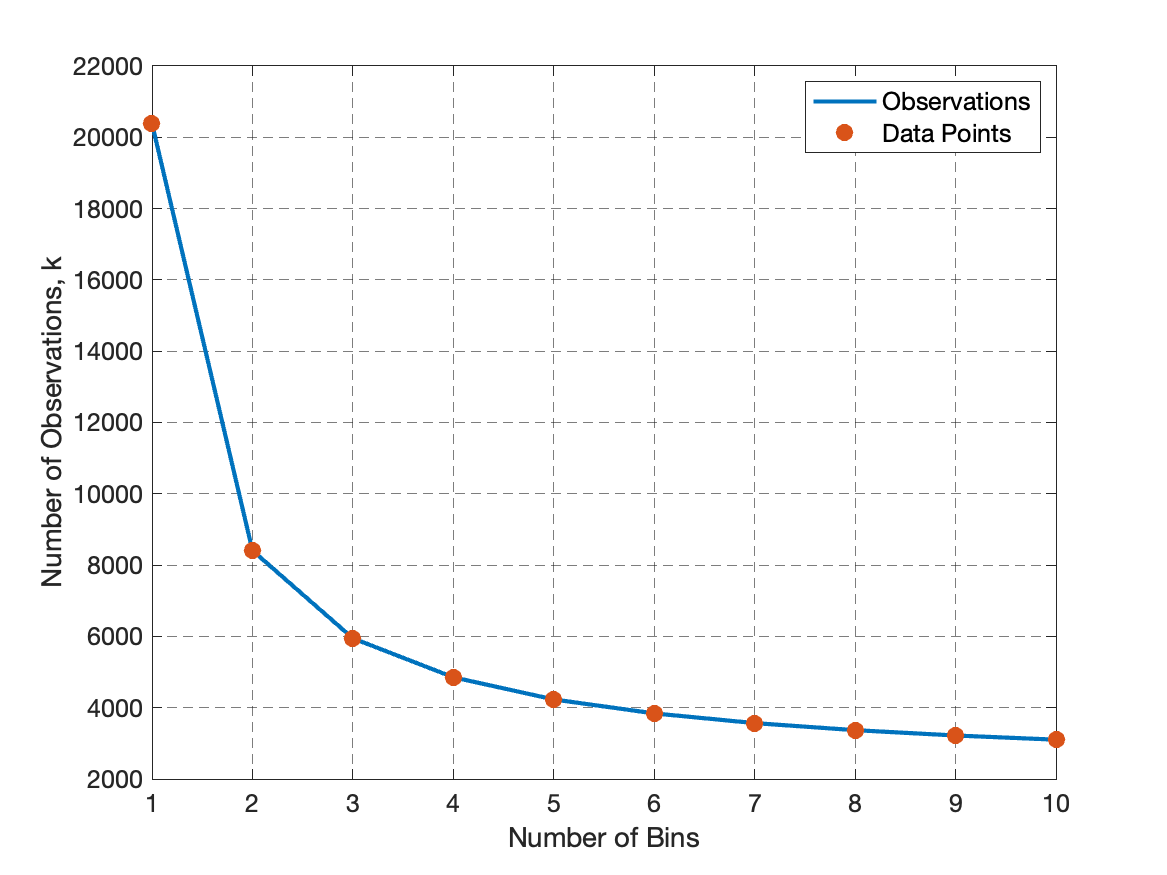}
    \caption{$h=1$.}
\end{subfigure}
\caption{The number of observations, $k$, needed to discriminate
  between $\vec{d}_t=(16.2,0.25,0.2)$ and
  $\vec{d}^*=(16.9,0.3,0.25)$ as a function of the number of bins
  recognised by the $\gamma$-detectors, for $h=0.5, 1$. Here we use the
  error rate $\delta = 0.05$ and prior probability $p_0 =
  0.5$. \label{kp}}
\end{figure*}

Figure \ref{kp}, with $h=0.2$, illustrates the required number of
observations as a function of $b$. From this plot, it's clear that as
$b$ increases, the value of $k$ tends to decrease.

\begin{figure*}[h!]
\centering
\begin{subfigure}{0.4\textwidth}
  \includegraphics[width=\textwidth]{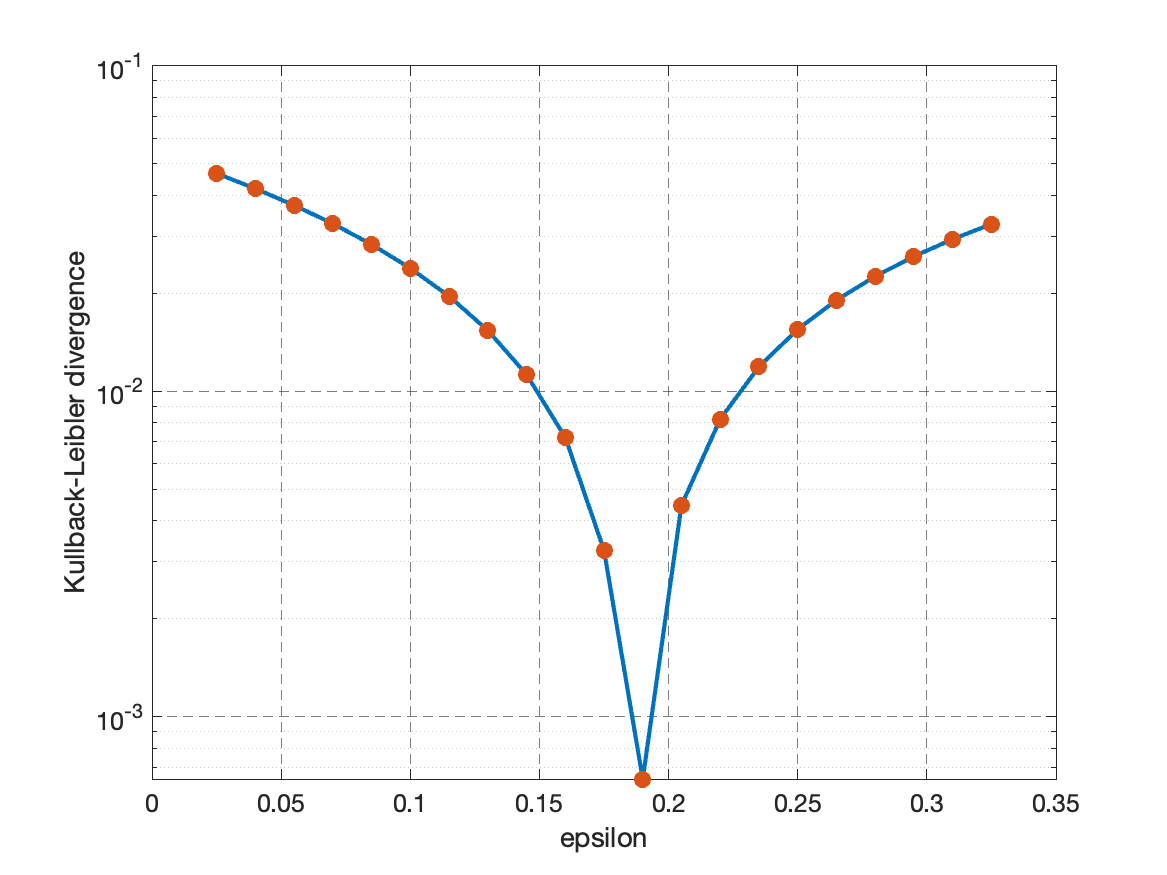}
    \caption{Varying $\epsilon$.}
\end{subfigure}
\begin{subfigure}{0.4\textwidth}
  \includegraphics[width=\textwidth]{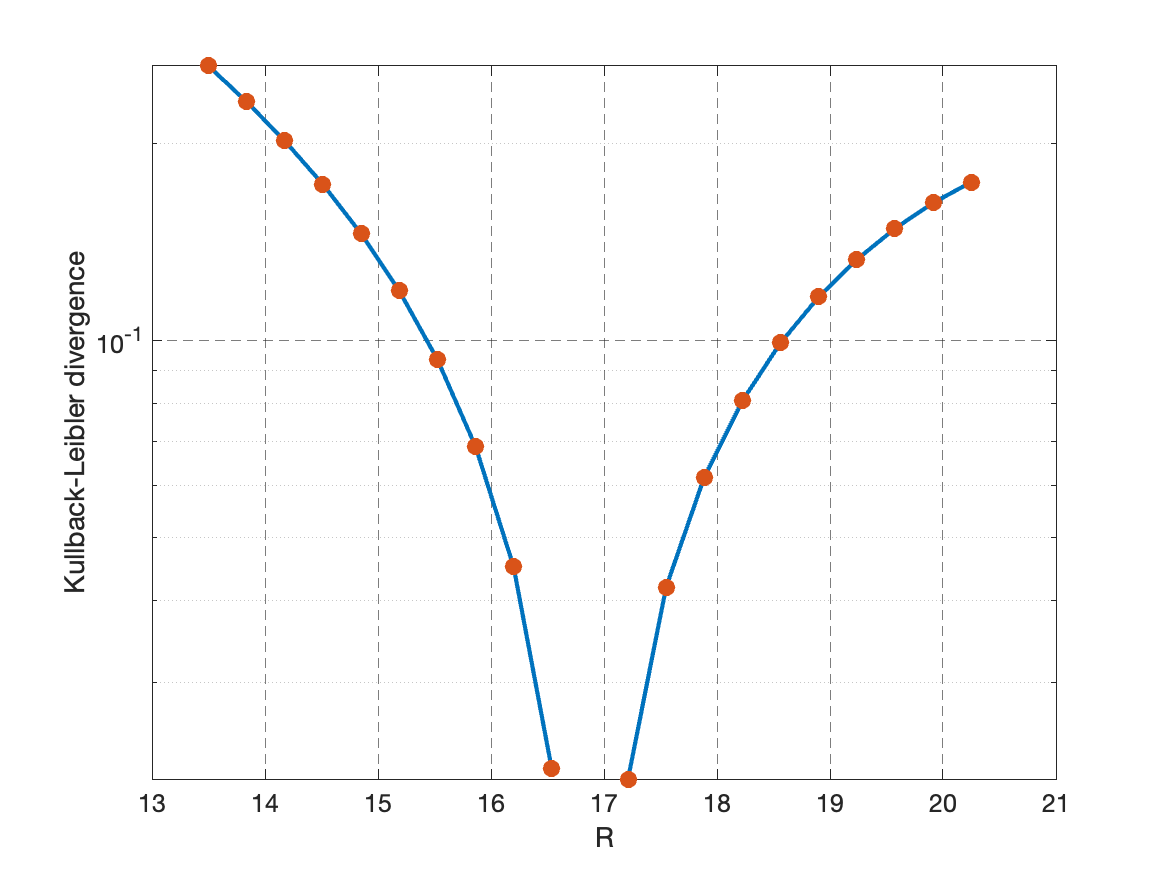}
    \caption{Varying $R$.}
\end{subfigure}
\begin{subfigure}{0.4\textwidth}
  \includegraphics[width=\textwidth]{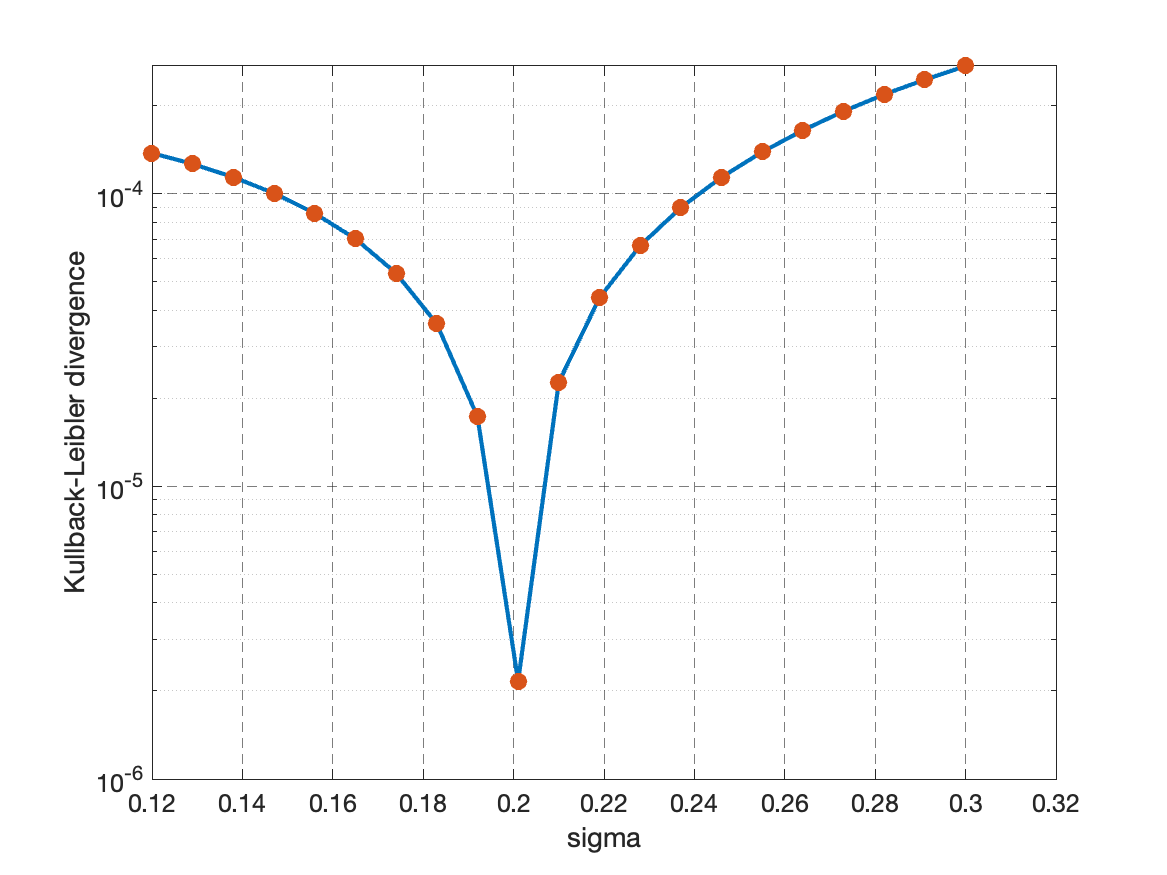}
    \caption{Varying $\sigma$.}
\end{subfigure}
\caption{ \label{fig:KL1d} Examination of the Kullback-Leibler
  divergence as a function of individual parameters for a true
  parameter range $\vec{d}=(16.9,0.3,0.25)$, a single detector and
  $h=1$. Notice that the dominant parameter is $R$.}
\end{figure*}

One can also examine the Kullback-Leibler divergence, $D_{KL}$, for
fixed $b$ and as a function of the parameter values, $\vec{d}$. This
enables examination of the sensitivity of the measure to the different
parametric values shown in Figure \ref{fig:KL1d}.

\section{Sequential Monte Carlo Approach}
\label{sec:SMC}

Sequential Monte Carlo (SMC) methods have been applied to a wide range
of Bayesian problems, and fundamentally rely on simulating a
collection of particles whose empirical measure is expected to follow
the posterior measure associated with a given data source. Their use
was popularised in e.g. \cite{Neal01,Doucet01,DelMoral04}. The problem
we face can be described as an IBIS (Iterated Batch Importance
Sampling) problem (\cite{Chopin02},
\cite[Section~17.2.2]{Chopin20}). Our numerical method will be based
on using a MCMC (Markov Chain Monte Carlo) step within an SMC
algorithm, e.g. \cite[Section~3.3.2.3]{DelMoral06}. To ensure that the
proposal distributions from the MCMC step remain viable, we introduce
an adaptive MCMC methodology, similar to that proposed in
\cite{Fearnhead13}.

Since it is difficult to explicitly calculate (\ref{bt}) due to the
term $p(\vec{x}')$, which would require integrating over space of
possible models, we can instead sample from the posterior distribution
using SMC. This involves having $N$ particles that each correspond to
some choice of $\vec{d}= \{ R_i, \sigma_i, \epsilon_i \}_{i=1}^n$,
sampled from the prior, which are then filtered using importance
weights defined by the likelihood function.  With additional
observations, the particles begin to cluster around the `true'
solution, representing a good approximation for $\vec{d}$. Below we
describe the algorithm as we have implemented it, including generating
data using a known model (`truth'). In a genuine implementation, this
would be replaced by the actual data observed by the monitoring
equipment.

Our SMC approach is detailed below:
\begin{enumerate}
\item \emph{Truth}: Select the `true' values for the model parameters $\vec d^*$, and then compute the corresponding `true' probability distribution of (\ref{dprobx}). 
   \item \emph{Initialisation}: For $i=1,2,\ldots,N$, sample $\vec{d}^i \sim p(\vec{d})$, where $p(\vec{d})$ is the Gaussian white noise prior.
   \item \emph{Sample Data}: Generate $k$ new observations by randomly
     sampling from the `true' probability distribution $p(\vec{x}|\vec{d}^*)$.
  \item \emph{Importance}: Given the new observations $x'_j,\;j=1,2,\ldots,k$, for $i=1,2,\ldots,N$, calculate the importance weights
  \begin{equation*}
    w_i=\prod_{j=1}^{k}P(x_j'|\vec{d}),\end{equation*}
  where $P(x_j'|\vec{d})$ is given by (\ref{dprobx}). Then
  normalise the weights to sum to one by dividing each weight by
  $\sum_{i=1}^n w_i$ to obtain $\widehat{w}_i=w_i/\sum_{i=1}^n w_i$. 
  \item \emph{Selection}: From the set $\{\vec{d}^1,\dots,
    \vec{d}^N\}$, resample with replacement $N$ particles
    $\{\widehat{\vec{d}}^1, \dots, \widehat{\vec{d}}^N\}$ with
    probabilities $\widehat{w}_1,\dots, \widehat{w}_N$.
  \item \emph{Mutation}: For $i=1, \dots, N$, slightly shift
    $\widehat{\vec{d}}^i$ in state space. See below for further
    detail about this step. These perturbed particles are set to
    $\vec{d}^i$.
  \item  \emph{Repeat}: While there is more data, go to step $(iii)$.
\end{enumerate}

To correctly perturb the models in the mutation step, we use a version of a Metropolis-Hastings algorithm, see e.g. \cite{Fearnhead13}, detailed below:
\begin{enumerate}
  \item For $i=1,2,\ldots,N$, sample $\vec{y}^i\sim N(\widehat{\vec{d}}^i,C_i)$, where $C_i$ is the covariance matrix for the proposal distribution (a Gaussian distribution has been chosen here).
  \item For $i=1,2,\ldots,N$, calculate the acceptance probability
  \begin{equation*}
  \alpha(\vec{y}^i|\widehat{\vec{d}}^i)=\min\left\{1,\frac{p(\vec{y}^i)\prod_{j=1}^{k}P(x_j'|\vec{y}^i)}{p(\widehat{\vec{d}}^i)\prod_{j=1}^{k}P(x_j'|\widehat{\vec{d}}^i)}\right\}.
  \end{equation*}
  This choice of $\alpha$ ensures that the more likely state according to the posterior is selected.
  \item For $i=1,2,\ldots,N$, sample $\vec{d}^i$ from $\{\widehat{\vec{d}}^i,\vec{y}^i\}$ with probability $\alpha$.
\end{enumerate}

Haario et al. (\cite{Haario01}) proposed a method for computing the covariance matrix such that it varies according to the particle's path, which they labelled Adaptive Metropolis. Here, to represent the history of the particle, we use the subscript $t$ to correspond to the number of iterations performed. The covariance matrix at iteration $t$ for particle $i$ is defined as
\begin{equation}
C_{i,t}=s_d \text{cov}(\widehat{\vec{d}}_{i,0},\ldots,\widehat{\vec{d}}_{i,t-1})+s_d\delta I_d,\end{equation}
where $s_d=(2.38^2)/m$, $m$ is the dimension of the vector $\widehat{\vec{d}}_i$, $I_d$ is the identity matrix and $\delta=0.1$ ensures that $C_{i,t}$ remains positive semi-definite. Note that we only consider the location of the particle after the importance sampling at each iteration. The definition of the covariance matrix for the points $\vec{X}_{0},\ldots,\vec{X}_{k}\in\vec{R}^d$ in state space is
\begin{equation*}
\text{cov}(\vec{X}_{0},\ldots,\vec{X}_{k})=\frac{1}{k}\left(\sum_{j=0}^k \vec{X}_{j}\vec{X}_{j}^T-(k+1)\bar{\vec{X}}\bar{\vec{X}}^T\right),
\end{equation*}
where $\bar{\vec{X}}=(1/(k+1))\sum_{j=0}^k\vec{X}_j$.

Roberts \& Rosenthal (\cite{Roberts09}) built upon this work by outlining different extensions to the Adaptive Metropolis technique. One approach involved having two phases for resampling such that
\begin{align*}
&\vec{y}^i\sim N\left(\widehat{\vec{d}}^i,\frac{0.01I_d}{d}\right),\quad \text{for }i\le 2d,\\
&\vec{y}^i\sim (1-\beta)N\left(\widehat{\vec{d}}^i,C_{i,t}\right)+\beta N\left(\widehat{\vec{m}}^i,\frac{0.01I_d}{d}\right),\quad \text{for }i> 2d,
\end{align*}
where $\beta=0.05$. The initial phase ensures that mutations do occur during `burn-in'. Alternative methods are described in \cite{Fearnhead13}, in particular including only implementing the mutation step in an adaptive manner, and may be necessary to reduce computational complexity. Note that we would anticipate that increased dimensionality may also prove challenging (see the discussion in \cite[Section~17.2.2]{Chopin20}), but could be addressed through the addition of tempering in the algorithm.

\section{Numerical results}
\label{sec:numerics}

To highlight the performance and showcase the main ideas behind our
methodology, we give a series of numerical examples to test our
algorithm. These are designed to investigate the effect of increasing
the number of bins, $b$, and the ability to infer tissue density and
range verification based on the \pg observations.

The accuracy is measured using the mean KL-divergence of the
particles,
\begin{align*}
  \widehat{KL}
  =
  \frac{1}{N}
  \sum_{j=1}^N D_{KL}(P(x'|\vec{d}^j)|P(x'|\vec{d}^*).
\end{align*}
Here, we always perform twenty iterations of the SMC algorithm, each
incorporating additional data, and then assess the prediction accuracy
at this point. As a result, all the simulations presented will have
roughly consistent computation times. 

\subsection{Range uncertanties of a pristine Bragg Peak}
\label{sec:PBP}

\begin{figure*}[ht!]
\begin{subfigure}{0.45\textwidth}
  \includegraphics[width=\textwidth]{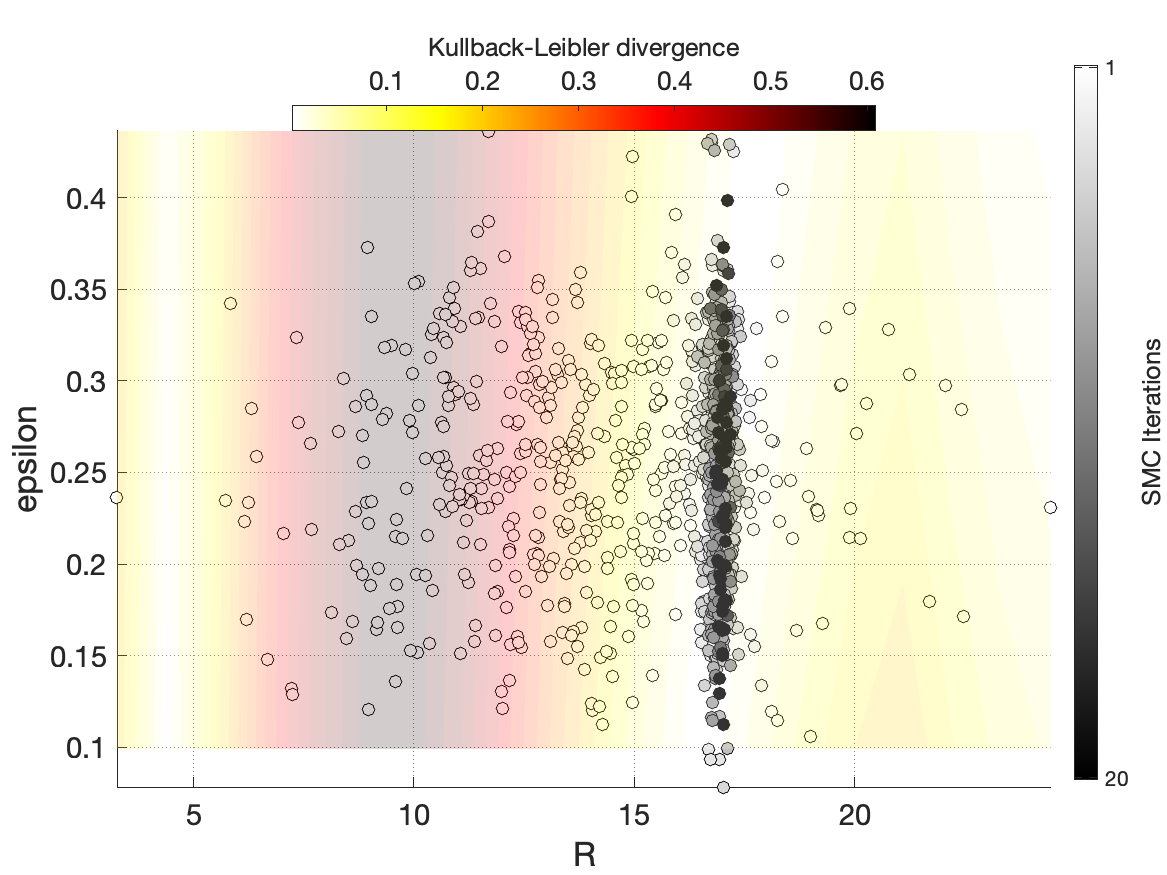}
    \caption{$R$-$\epsilon$ space with a single detector.}
\end{subfigure}
\hfill
\begin{subfigure}{0.45\textwidth}
  \includegraphics[width=\textwidth]{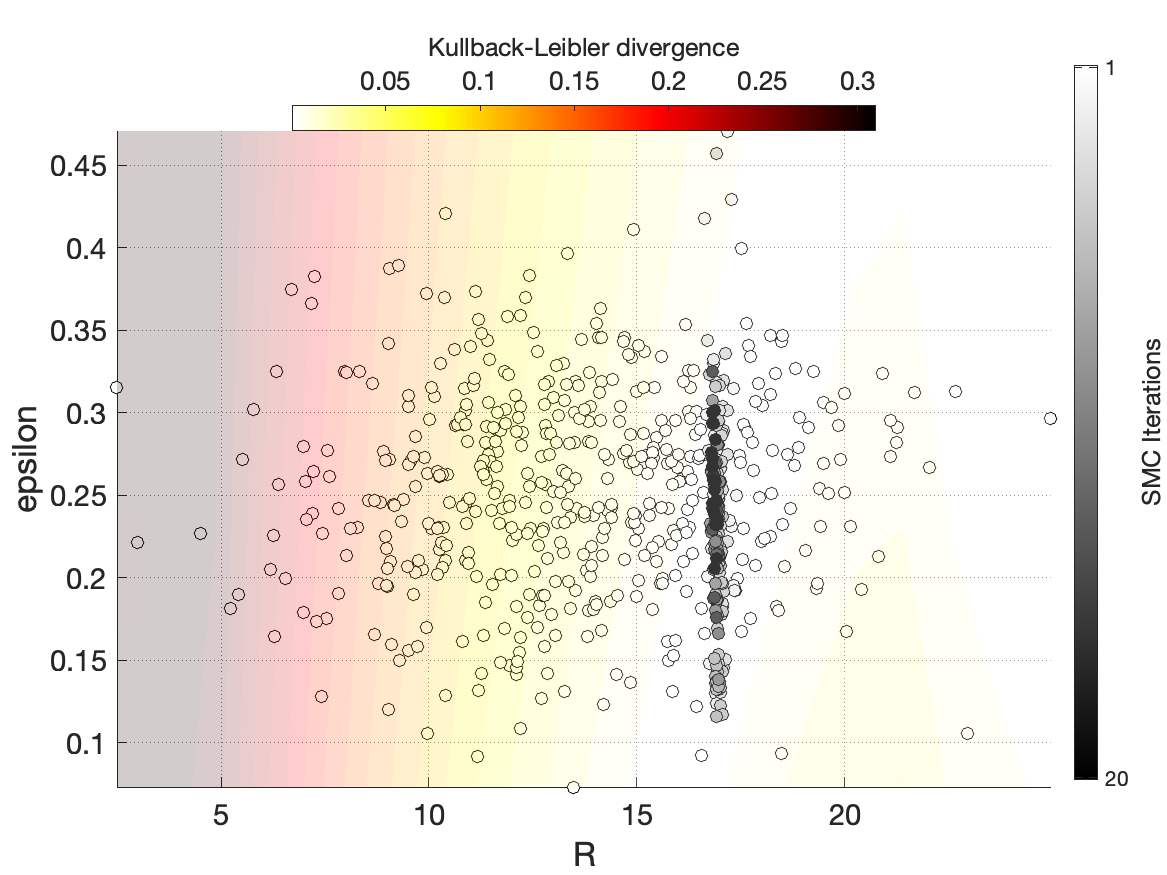}
    \caption{$R$-$\epsilon$ space with six detectors.}
\end{subfigure}
\\
\begin{subfigure}{0.45\textwidth}
  \includegraphics[width=\textwidth]{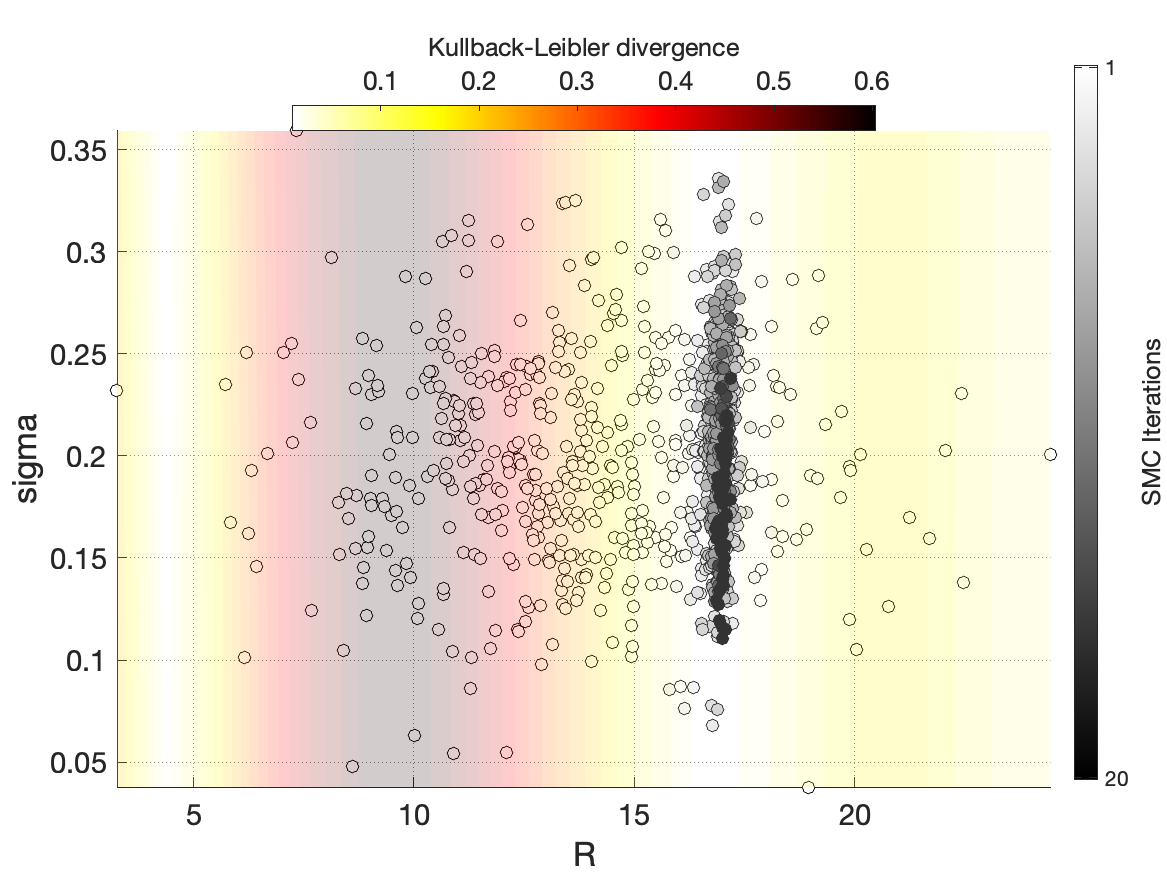}
    \caption{$R$-$\sigma$ space with a single detector.}
\end{subfigure}
\hfill
\begin{subfigure}{0.45\textwidth}
  \includegraphics[width=\textwidth]{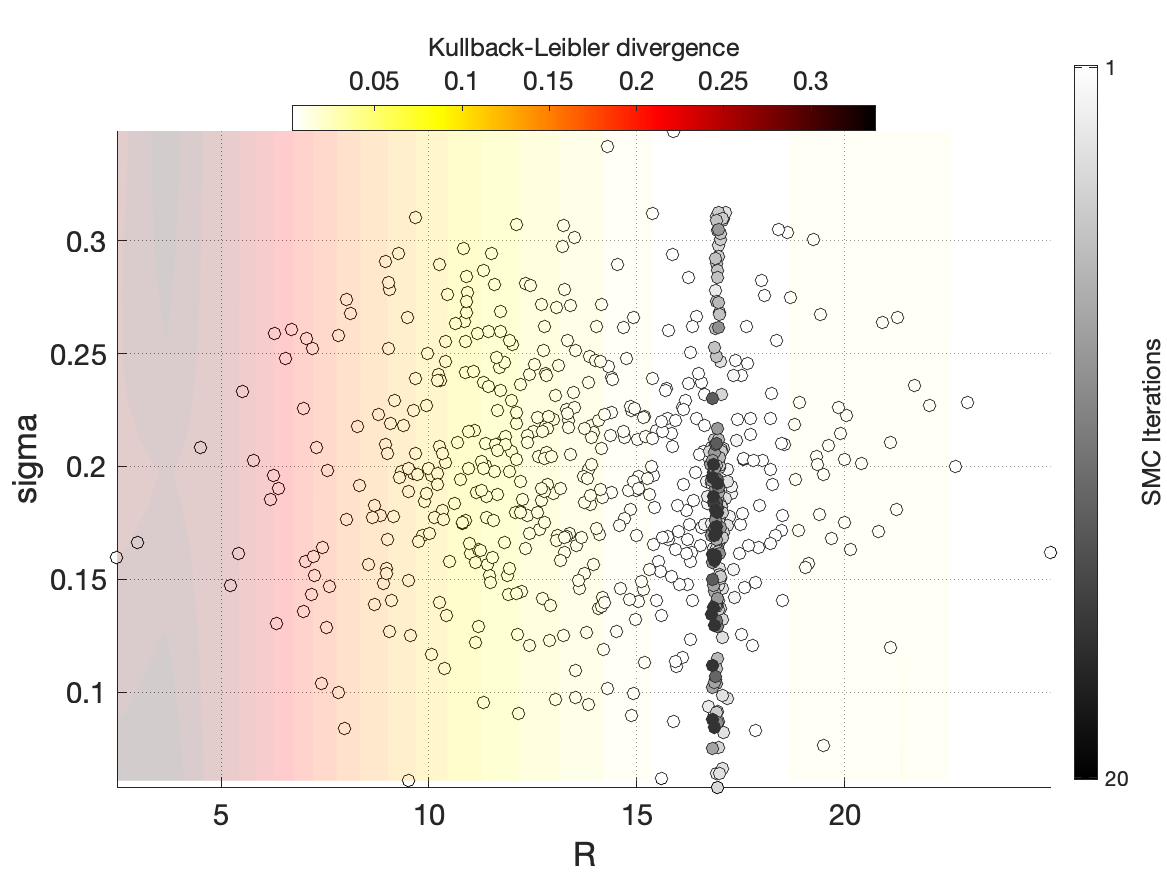}
    \caption{$R$-$\sigma$ space with  six detectors.}
\end{subfigure}
\\
\begin{subfigure}{0.45\textwidth}
  \includegraphics[width=\textwidth]{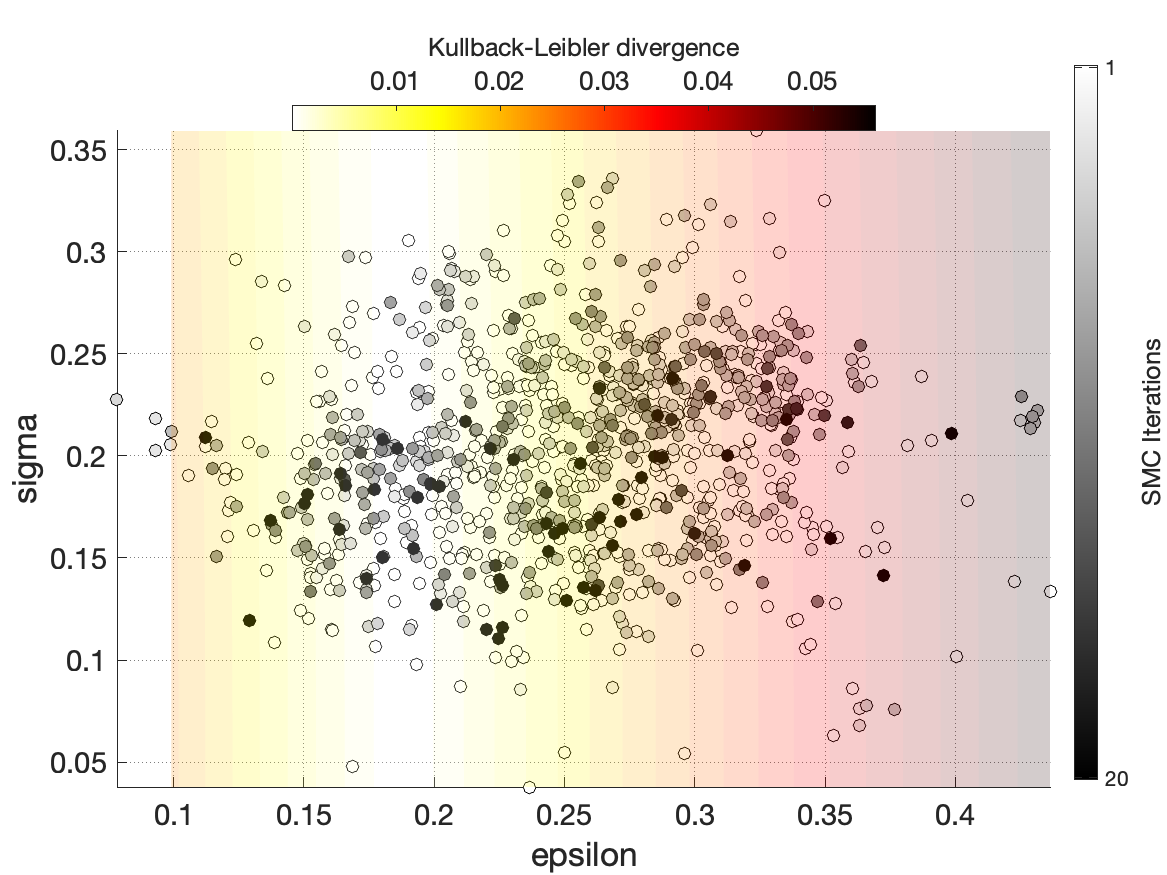}
    \caption{$\epsilon$-$\sigma$ space with a single detector.}
\end{subfigure}
\hfill
\begin{subfigure}{0.45\textwidth}
  \includegraphics[width=\textwidth]{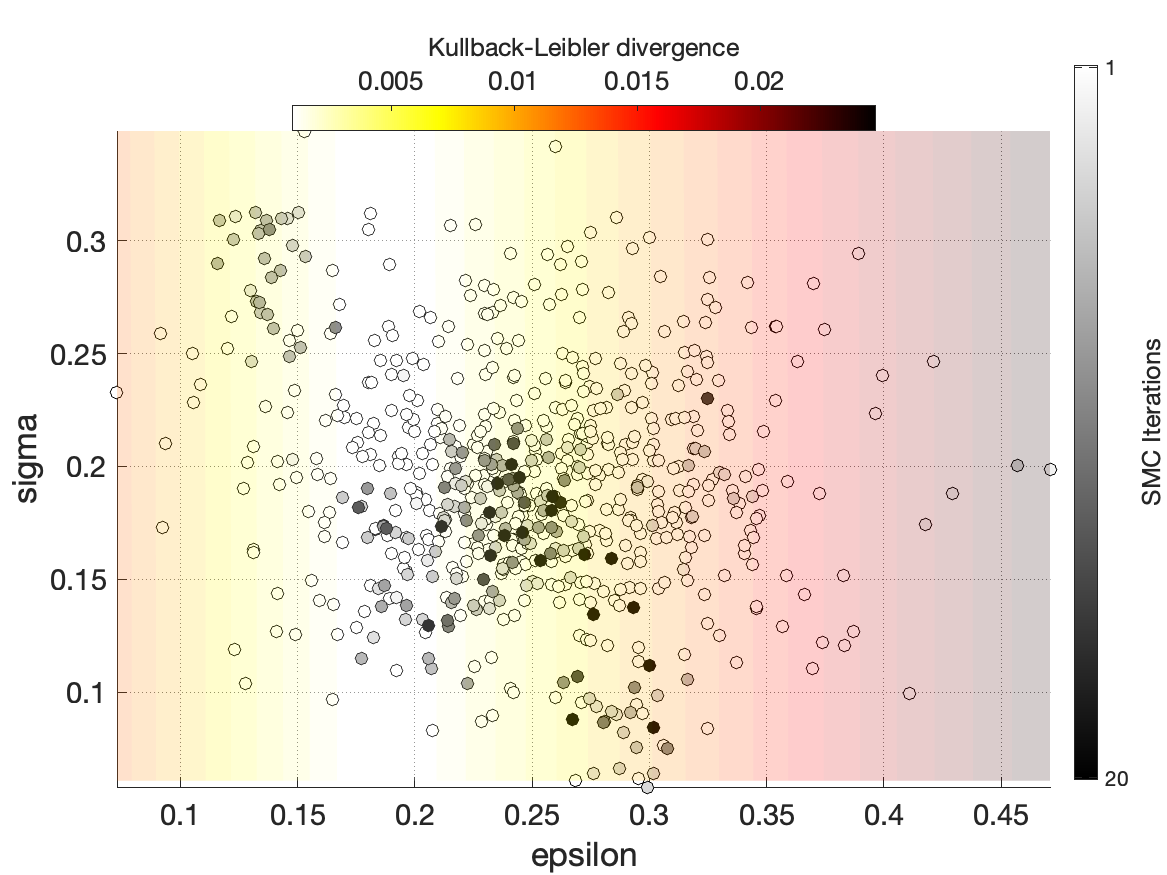}
    \caption{$\epsilon$-$\sigma$ space with  six detectors.}
\end{subfigure}
\caption{
  \label{fig:PBP1B}
  Numerical simulation of the SMC algorithm. The points correspond to
  particles representing parameter values, which concentrate as more
  data accumulates. Note that the particles are able to clearly
  identify the range accurately, but are not able to fully resolve
  uncertainty in the $\sigma, \epsilon$ parameters. The background
  denotes the KL divergence of the given points relative to the
  truth. Based on the given scales, it is clear that it is harder for
  the particles to resolve uncertainty in these parameters against the
  uncertainty in the range parameter.}
\end{figure*}

\begin{figure*}
\centering
\begin{subfigure}{0.4\textwidth}
  \includegraphics[width=\textwidth]{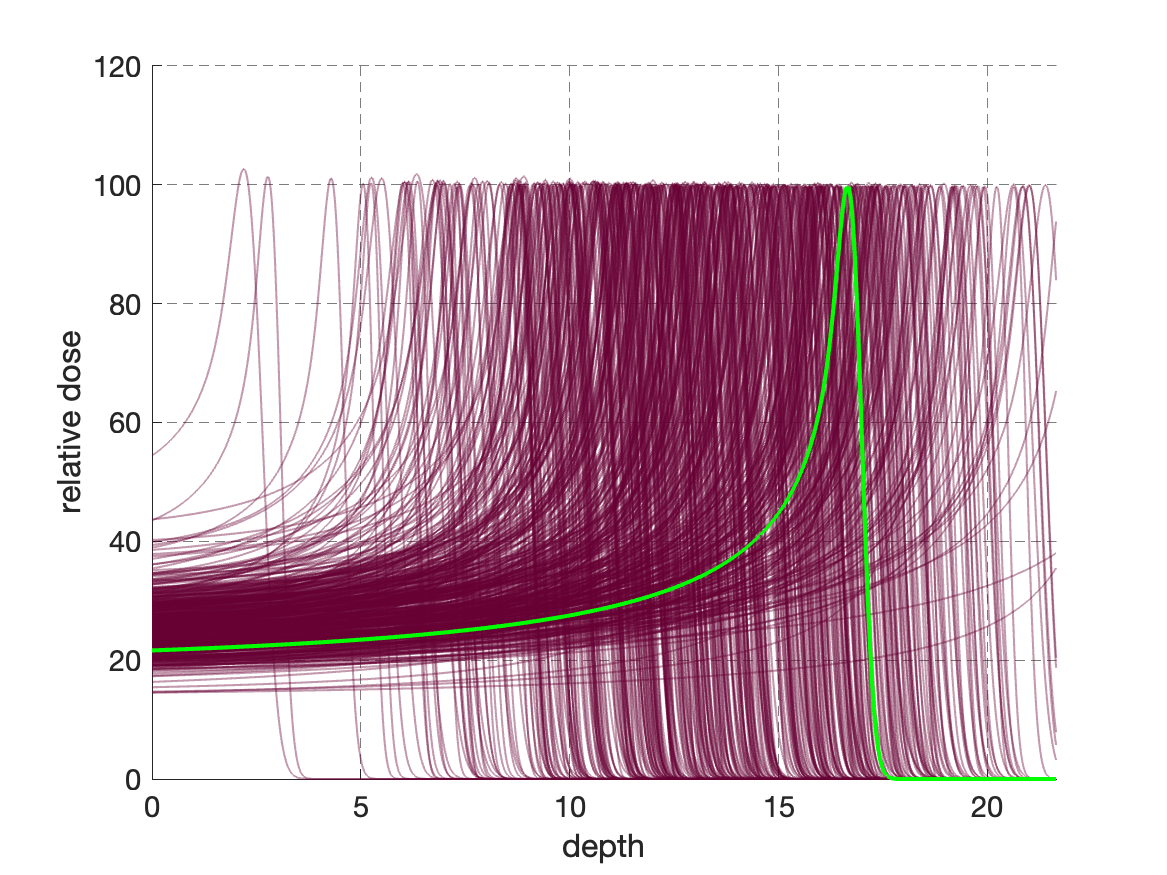}
    \caption{Posterior dose profiles ($Q$) after one SMC iteration.}
\end{subfigure}
\hfill
\begin{subfigure}{0.4\textwidth}
  \includegraphics[width=\textwidth]{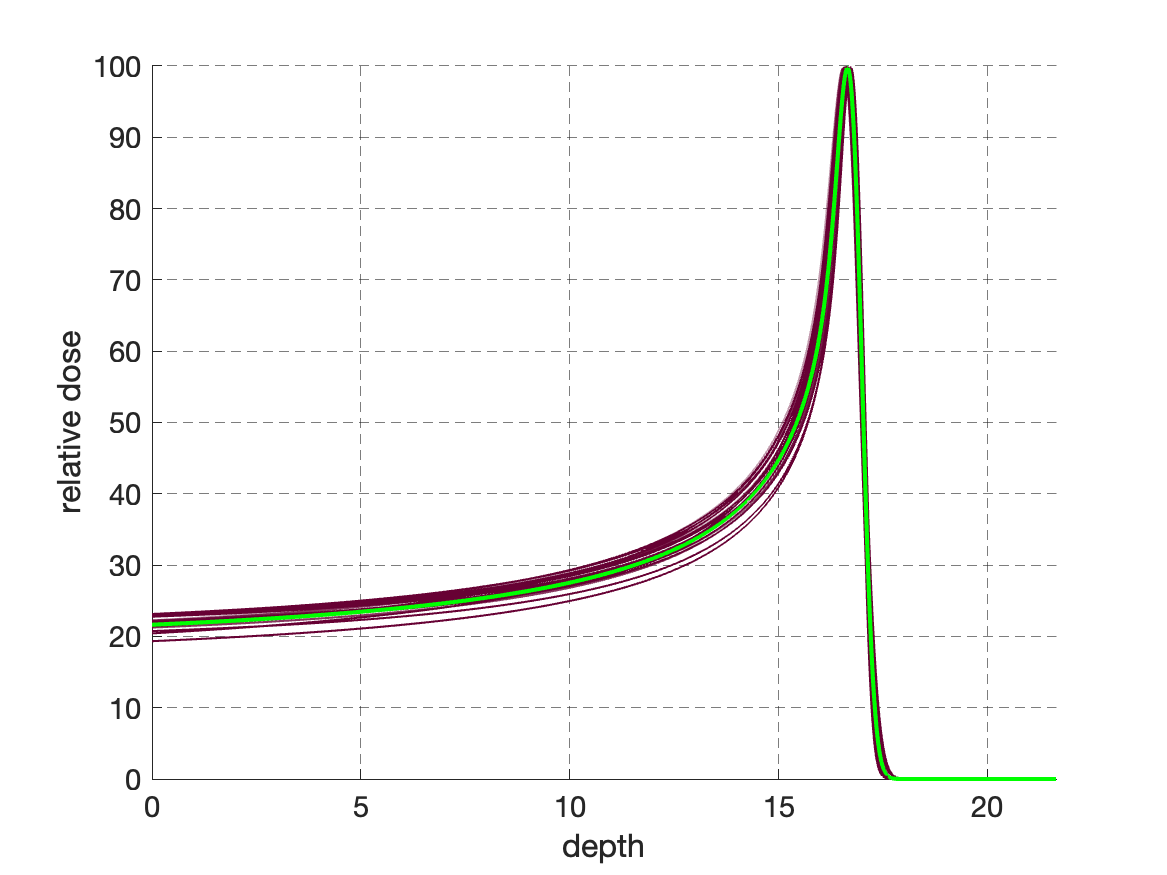}
    \caption{Posterior dose profiles ($Q$) after twenty SMC iterations.}
\end{subfigure}
\\
\begin{subfigure}{0.4\textwidth}
  \includegraphics[width=\textwidth]{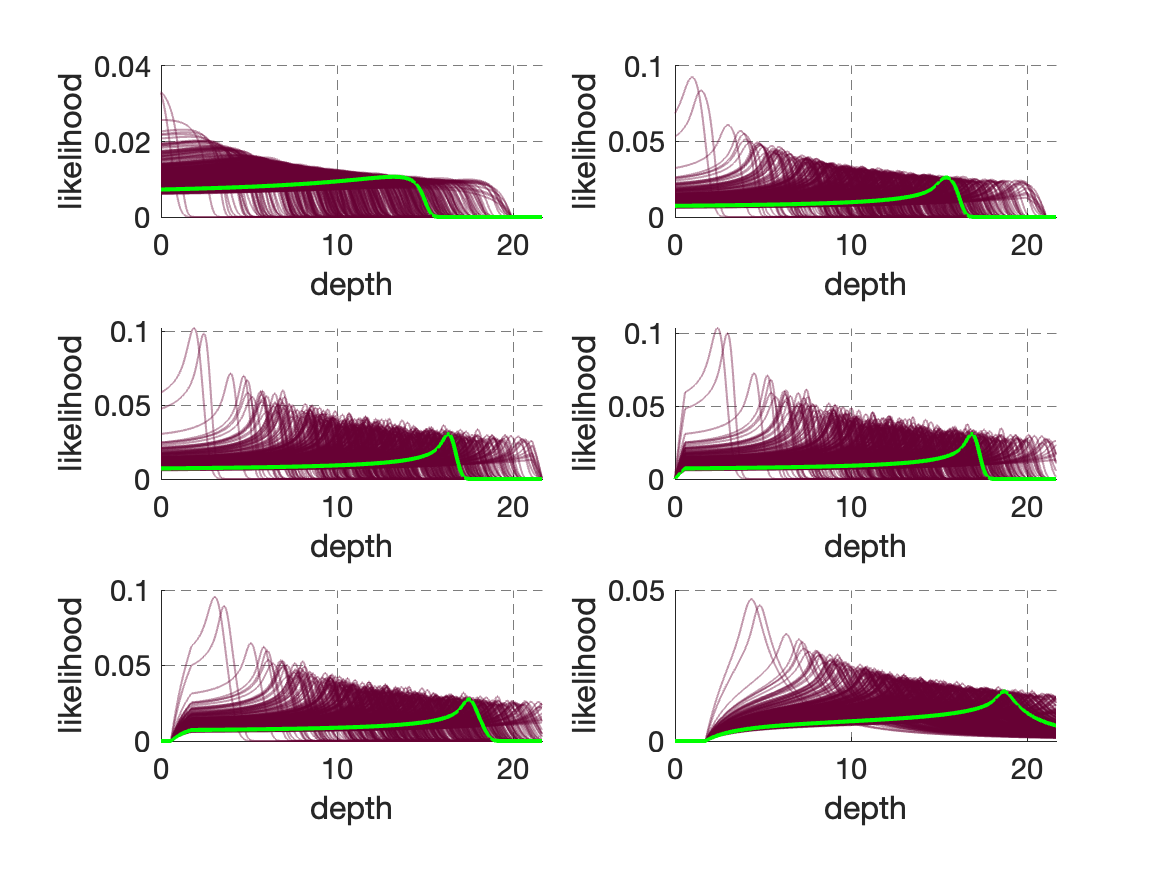}
    \caption{Posterior likelihood functions ($P$) after one SMC iteration.}
\end{subfigure}
\hfill
\begin{subfigure}{0.4\textwidth}
  \includegraphics[width=\textwidth]{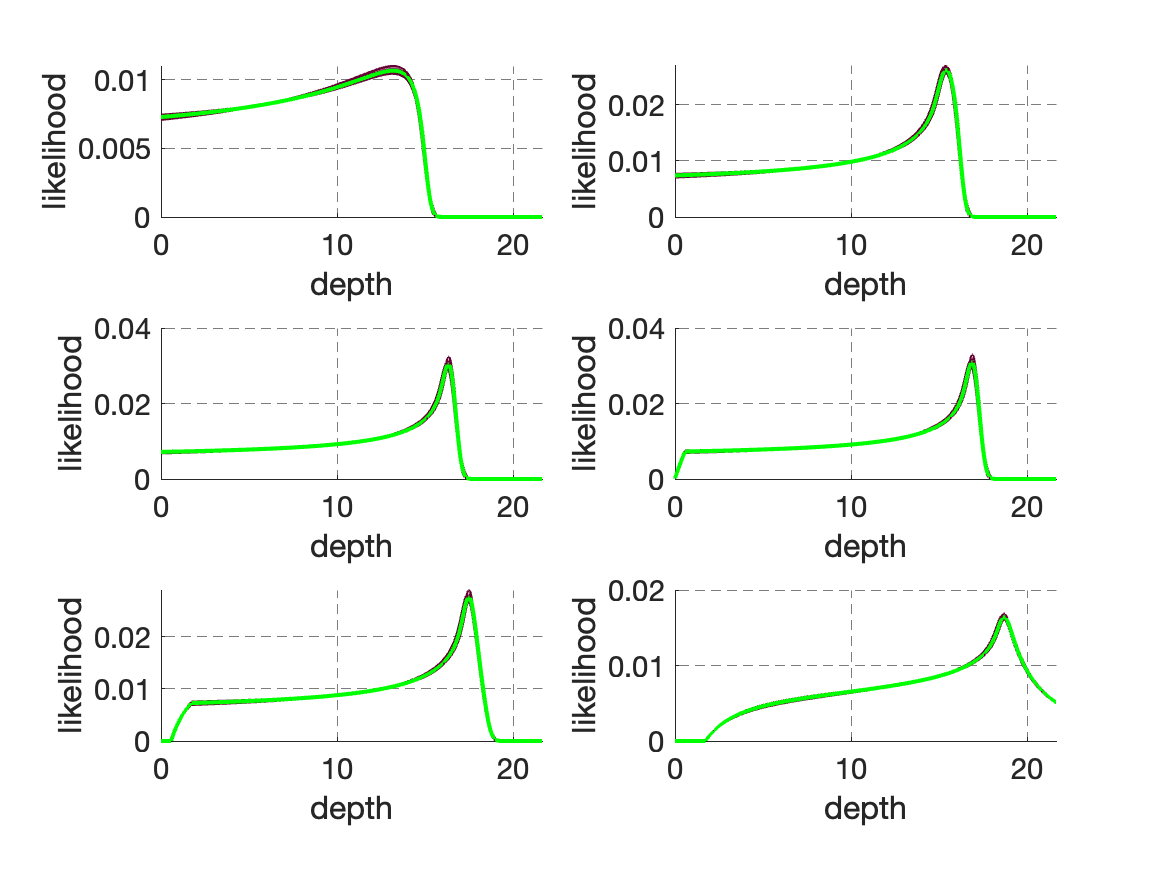}
    \caption{Posterior likelihood functions ($P$) after twenty SMC iterations.}
\end{subfigure}
\\
\begin{subfigure}{0.4\textwidth}
  \includegraphics[width=\textwidth]{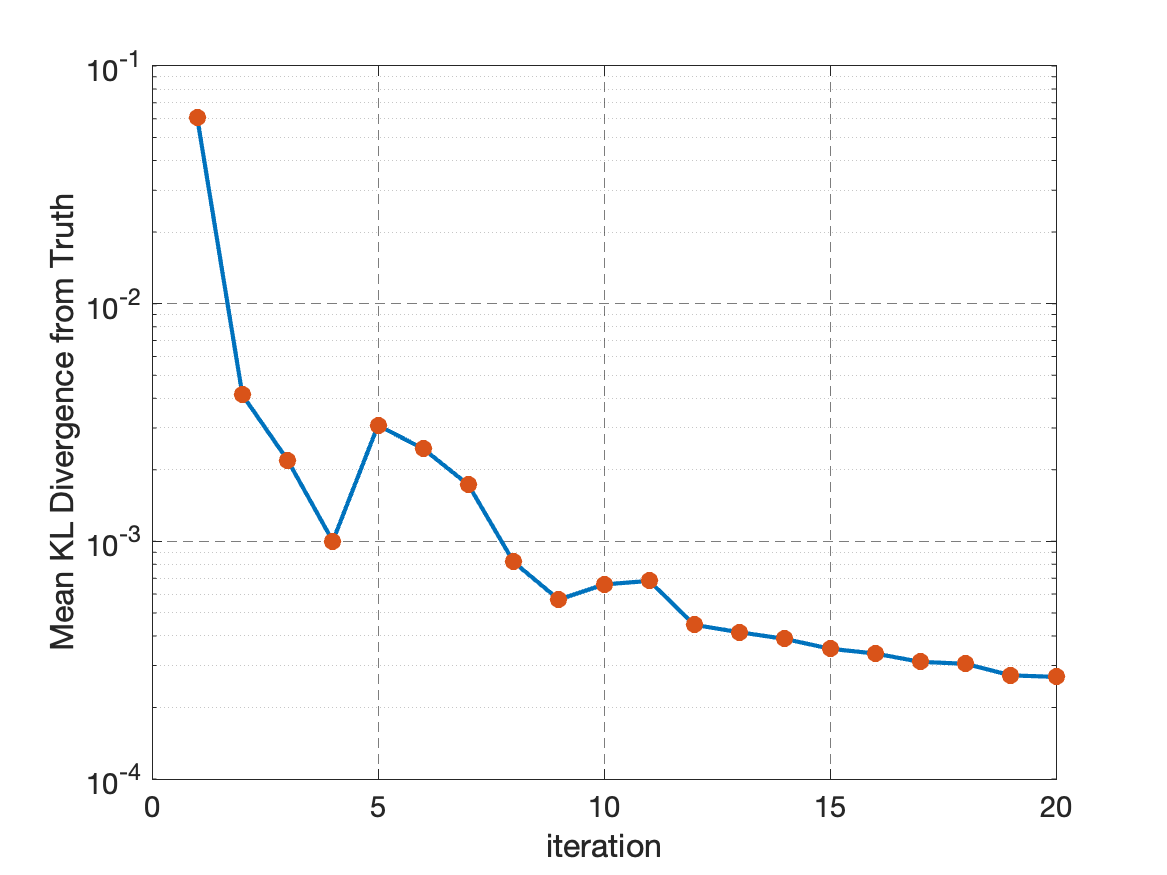}
  \caption{Kullback-Leibler divergence error as a function of SMC iterations.}
\end{subfigure}
\hfill
\begin{subfigure}{0.4\textwidth}
  \includegraphics[width=\textwidth]{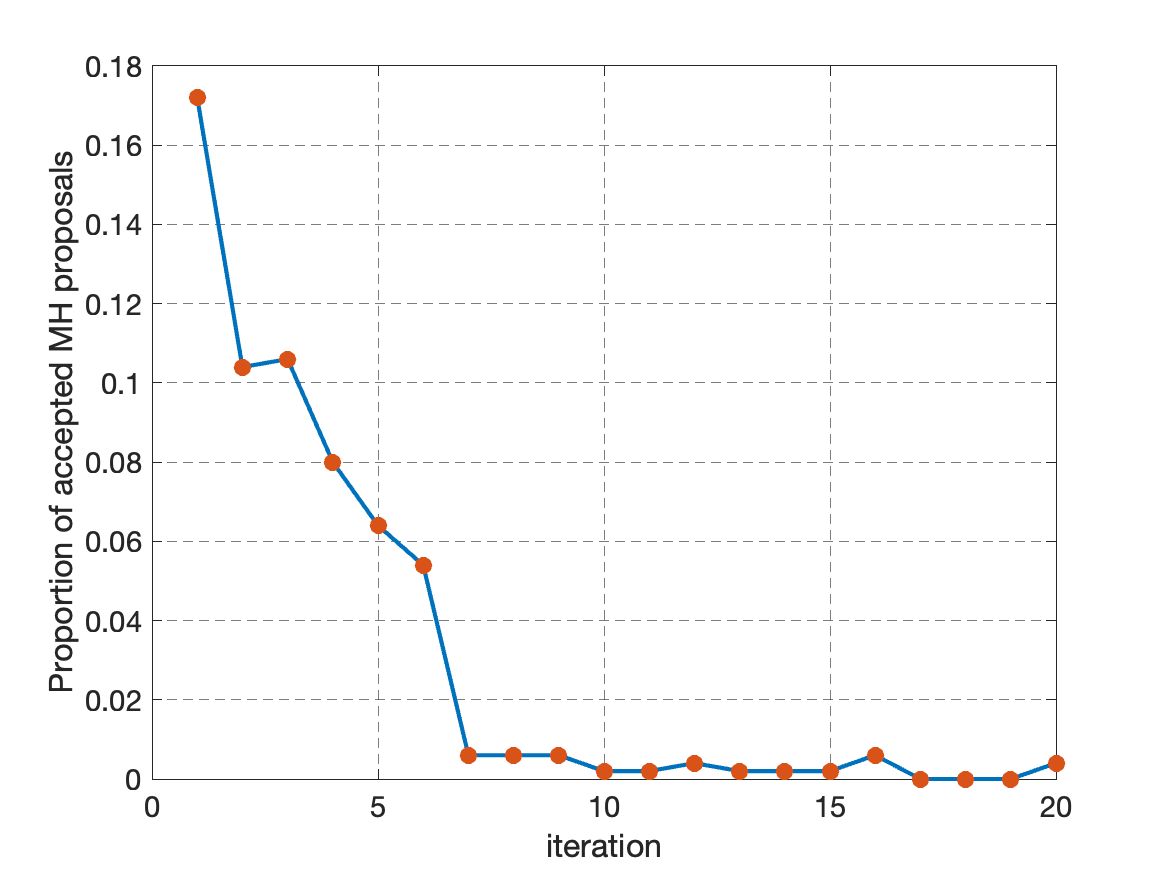}
  \caption{Proportion of accepted Metropolis Hastings proposals as a
    function of SMC iterations.}
\end{subfigure}
\caption{
  \label{fig:PBP6B}
  This Figure shows the experiment from Figure \label{fig:PBP1} with
  the exception that there are six detectors. Notice the algorithm
  converges significantly quicker and the realisation of dose profile
  at the twentieth iteration is more accurate than with a single
  detector.}
\end{figure*}

To begin we consider the case of a pristine Bragg Peak at an energy
level of 100MeV through a water phantom. In this case the Bortfeld
dose function can be written explicitly with three unknown parametric
values, $R, \sigma$ and $\epsilon$. The range of the beam, $R$, is
particularly pertinent as it is one of the key uncertainties in proton
therapy. We showcase the results of the algorithm presented in Figures
\ref{fig:PBP1B}--\ref{fig:PBP6B} where the problem is examined with a
single detector and six separate detectors respectively. It is worth
noting that the algorithm converges significantly quicker when using
multiple detectors, although even with a single one the dose is
reproduced accurately after twenty SMC iterations.

\begin{figure*}[ht!]
  \centering
  \begin{subfigure}{0.4\textwidth}
    \includegraphics[width=\textwidth]{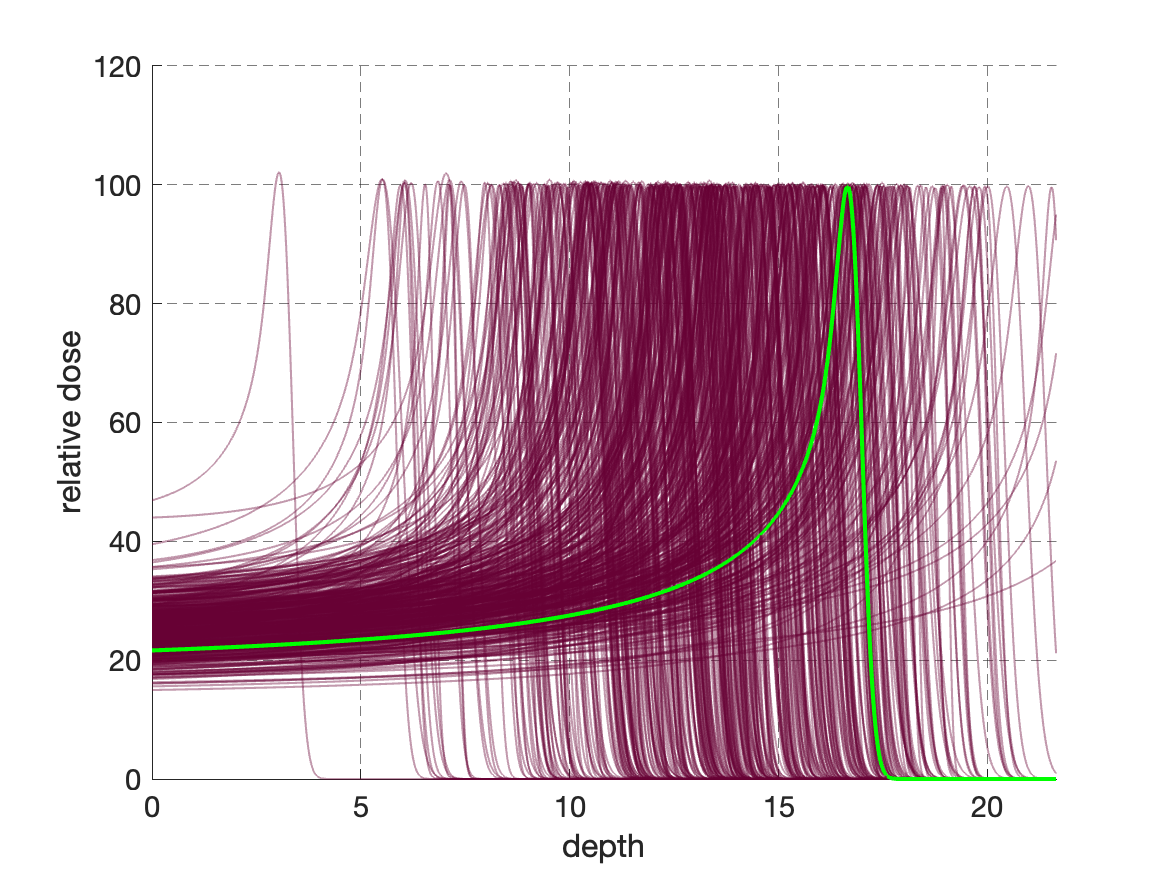}
    \caption{Posterior dose profiles ($Q$) after one SMC iteration.}
  \end{subfigure}
  \hfill
  \begin{subfigure}{0.4\textwidth}
    \includegraphics[width=\textwidth]{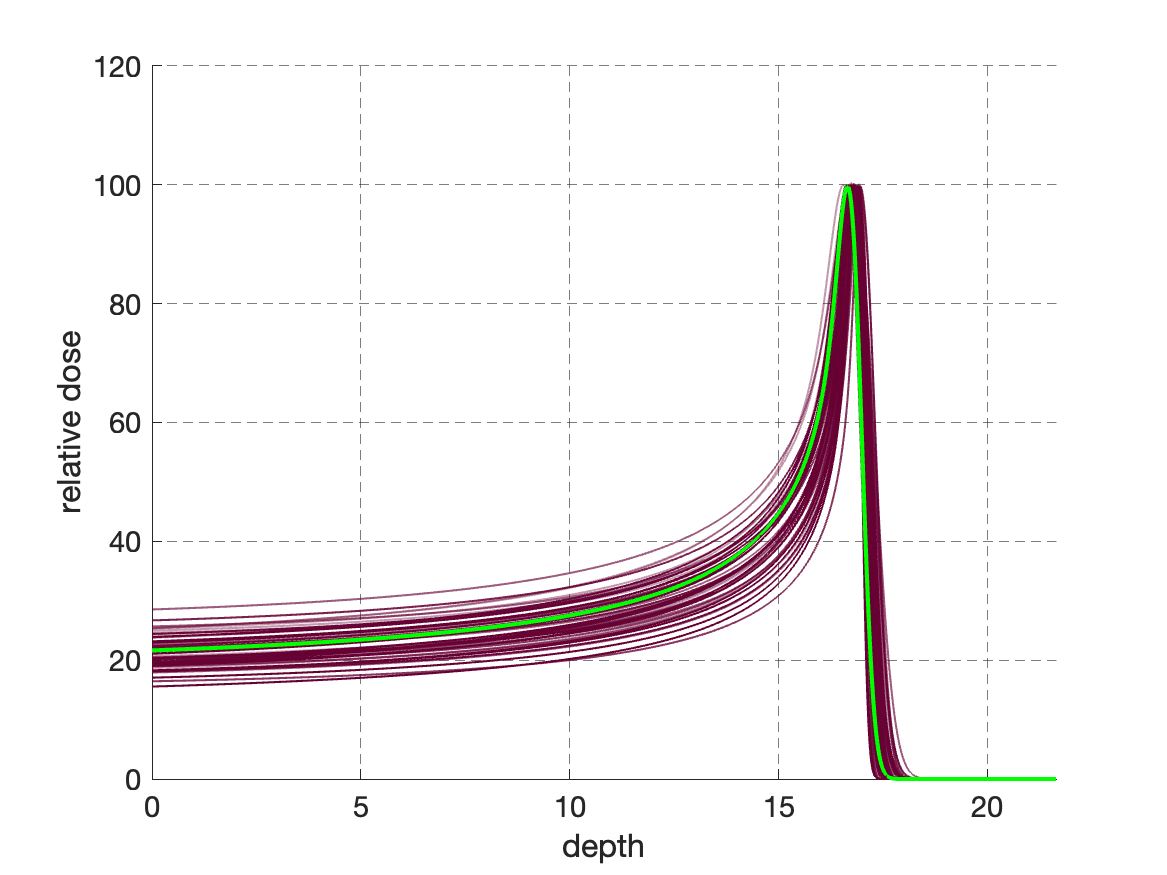}
    \caption{Posterior dose profiles ($Q$) after twenty SMC iterations.}
  \end{subfigure}
  \\
  \begin{subfigure}{0.4\textwidth}
    \includegraphics[width=\textwidth]{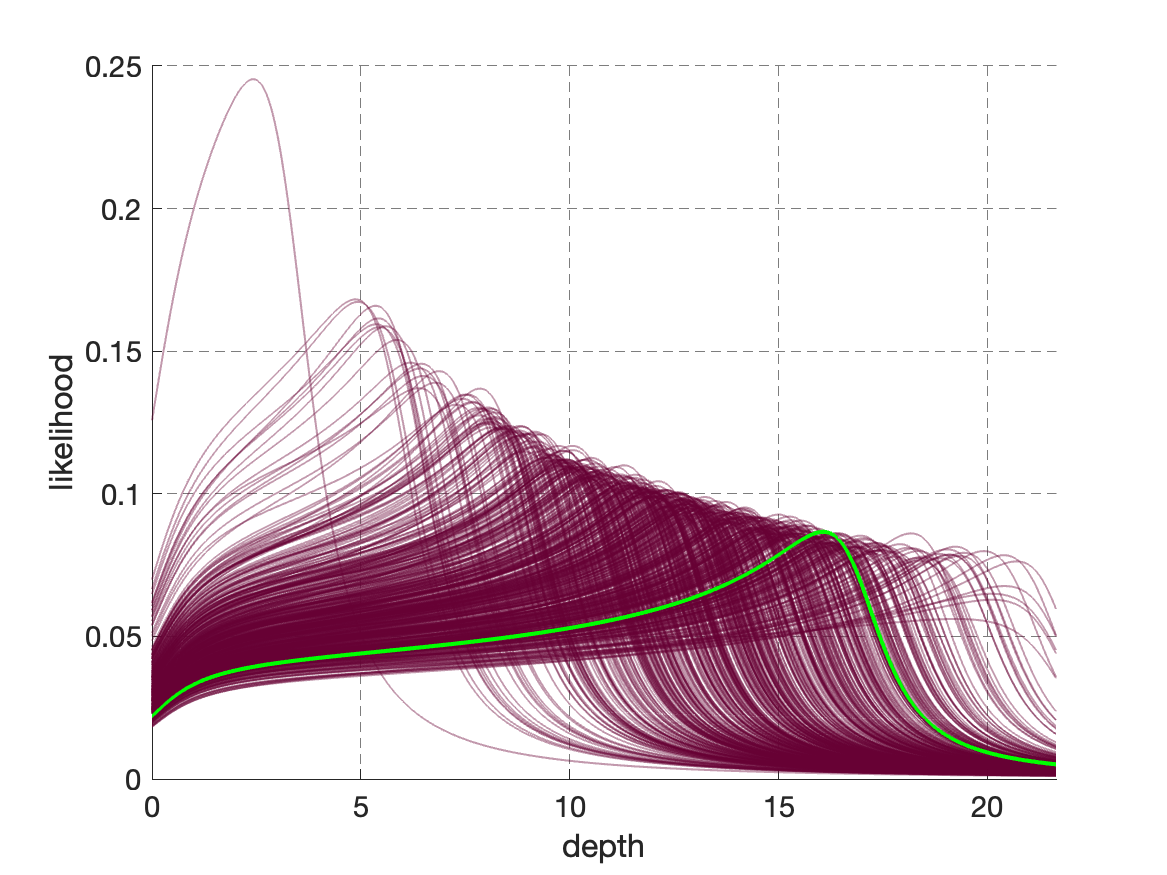}
    \caption{Posterior likelihood functions ($P$) after one SMC iteration.}
  \end{subfigure}
  \hfill
  \begin{subfigure}{0.4\textwidth}
    \includegraphics[width=\textwidth]{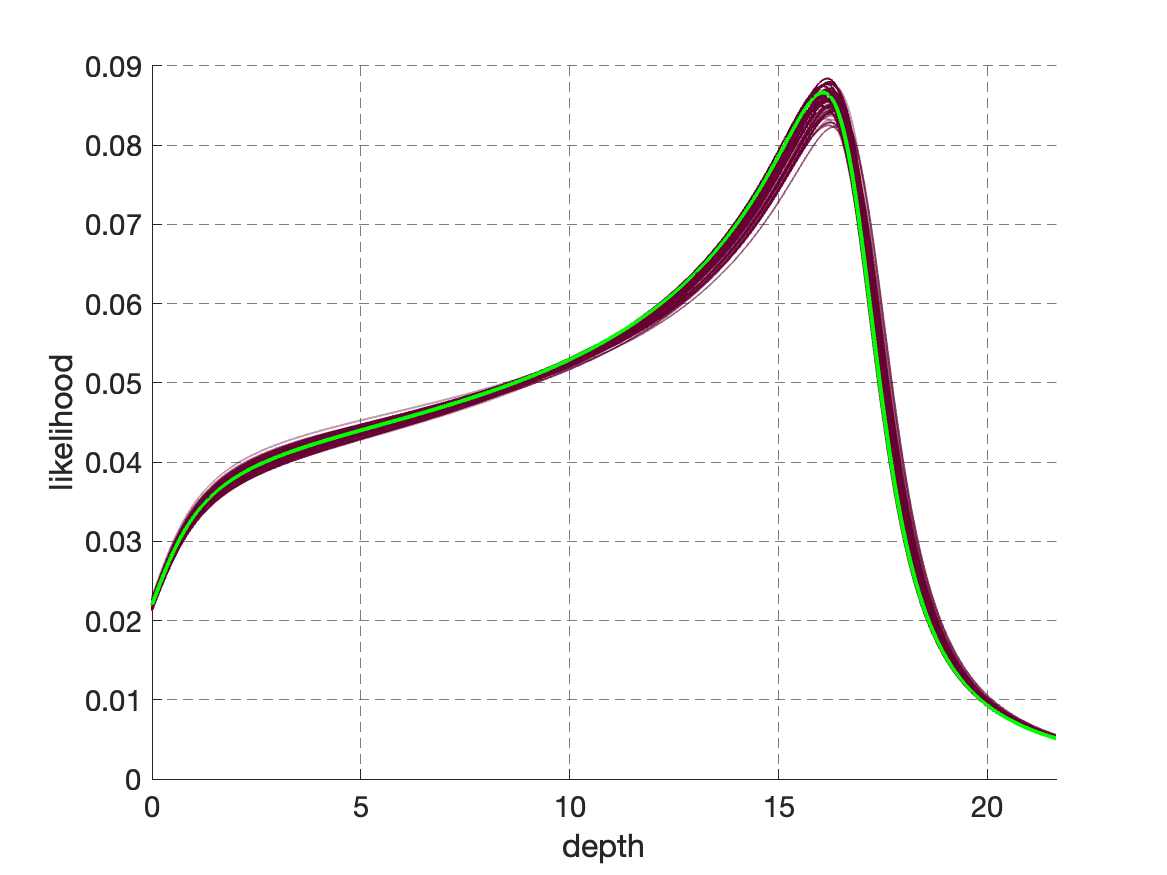}
    \caption{Posterior likelihood functions ($P$) after twenty SMC iterations.}
  \end{subfigure}
  \\
  \begin{subfigure}{0.4\textwidth}
    \includegraphics[width=\textwidth]{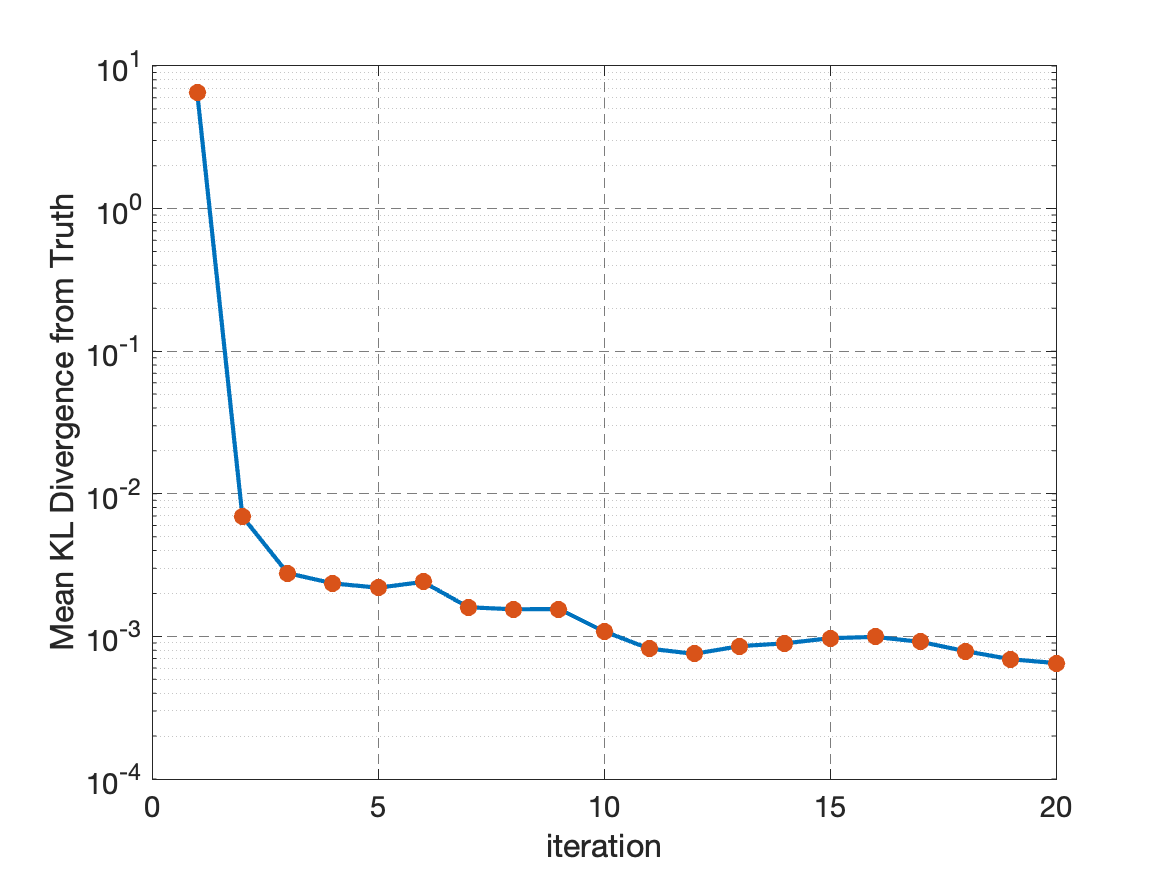}
    \caption{Kullback-Leibler divergence error as a function of SMC iterations.}
  \end{subfigure}
  \hfill
  \begin{subfigure}{0.4\textwidth}
    \includegraphics[width=\textwidth]{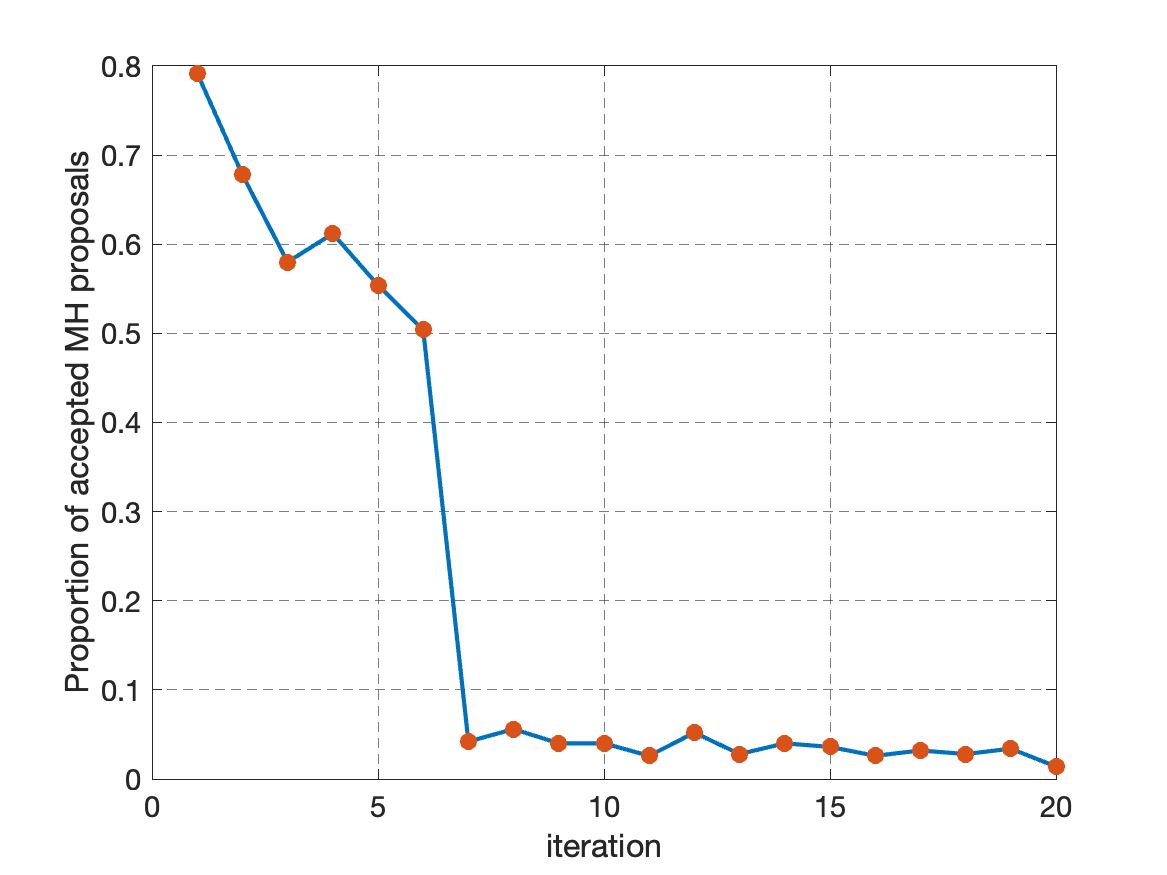}
    \caption{Proportion of accepted Metropolis Hastings proposals as a
      function of SMC iterations.}
  \end{subfigure}
  \label{fig:PBP1}
  \caption{The figures above show true and posterior samples of dose
    profiles (a,b) and observation (c,d) distributions. In (a,b), the
    green curve denotes the true dose profile,
    $Q(\cdot|\vec{d}^*)$, while the purple curves correspond to
    $Q(\cdot|\vec{d}_i)$ (c.f. (\ref{eq:Qdefn})) for samples from
    the posterior distribution. In (c,d), the curves correspond to the
    true (green) and posterior (purple) observation profiles,
    $P(\cdot|\vec{d}^*)$ and $P(\cdot|\vec{d}_i)$
    (c.f. (\ref{dprobx})) respectively. Plots (a,c) show the posterior
    distribution after one step of the SMC algorithm, corresponding to
    $k=1000$ observed $\gamma$-particles. Plots (b,d) correspond to 20
    steps of the SMC algorithm, with $10$ times as many
    observations. Distribution of the particles in
    $(R,\sigma,\epsilon)$ space is shown in Figure~\ref{fig:PBP1B}.}
\end{figure*}

\clearpage
\subsection{Lung tumour}
\label{sec:lung}

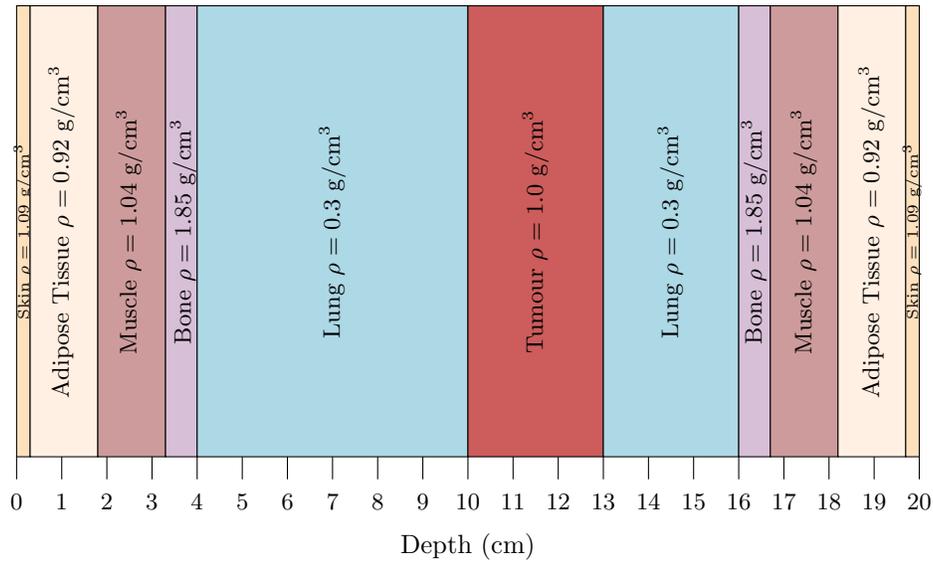
\begin{figure*}[h!]
  \centering
  \begin{tikzpicture}[scale=0.6]
    \definecolor{skinColor}{RGB}{255,224,189}
    \definecolor{adiposeColor}{RGB}{255,240,227}
    \definecolor{muscleColor}{RGB}{205,155,155}
    \definecolor{boneColor}{RGB}{216,191,216}
    \definecolor{lungColor}{RGB}{173,216,230}
    \definecolor{tumourColor}{RGB}{205,92,92}
    
    \draw[fill=skinColor] (0,0) rectangle (0.3,10);
    \node[rotate=90, font=\tiny] at (0.15,5) {Skin $\rho = 1.09$ g/cm$^3$};
    
    \draw[fill=adiposeColor] (0.3,0) rectangle (1.8,10);
    \node[rotate=90, font=\small] at (1,5) {Adipose Tissue $\rho = 0.92$ g/cm$^3$};
    
    \draw[fill=muscleColor] (1.8,0) rectangle (3.3,10);
    \node[rotate=90, font=\small] at (2.5,5) {Muscle $\rho = 1.04$ g/cm$^3$};
    
    \draw[fill=boneColor] (3.3,0) rectangle (4,10);
    \node[rotate=90, font=\small] at (3.7,5) {Bone $\rho = 1.85$ g/cm$^3$};
    
    \draw[fill=lungColor] (4,0) rectangle (10,10);
    \node[rotate=90, font=\small] at (7,5) {Lung $\rho = 0.3$ g/cm$^3$};
    
    \draw[fill=tumourColor] (10,0) rectangle (13,10);
    \node[rotate=90, font=\small] at (11.5,5) {Tumour $\rho = 1.0$ g/cm$^3$};
    
    \draw[fill=lungColor] (13,0) rectangle (16,10);
    \node[rotate=90, font=\small] at (14.5,5) {Lung $\rho = 0.3$ g/cm$^3$};
    
    \draw[fill=boneColor] (16,0) rectangle (16.7,10);
    \node[rotate=90, font=\small] at (16.35,5) {Bone $\rho = 1.85$ g/cm$^3$};
    
    \draw[fill=muscleColor] (16.7,0) rectangle (18.2,10);
    \node[rotate=90, font=\small] at (17.45,5) {Muscle $\rho = 1.04$ g/cm$^3$};
    
    \draw[fill=adiposeColor] (18.2,0) rectangle (19.7,10);
    \node[rotate=90, font=\small] at (18.95,5) {Adipose Tissue $\rho = 0.92$ g/cm$^3$};
    
    \draw[fill=skinColor] (19.7,0) rectangle (20,10);
    \node[rotate=90, font=\tiny] at (19.85,5) {Skin $\rho = 1.09$ g/cm$^3$};
    
    \node at (10,-2) {Depth (cm)};

    \foreach \x in {0,1,...,20} {
      \draw (\x,-0.5) -- (\x,0);
      \node at (\x, -1) {\small \x};
    }   
  \end{tikzpicture}
  \caption{\label{fig:lung} A one-dimensional lung cross section.}  
\end{figure*}

\begin{figure*}
\centering
\begin{subfigure}{0.4\textwidth}
  \includegraphics[width=\textwidth]{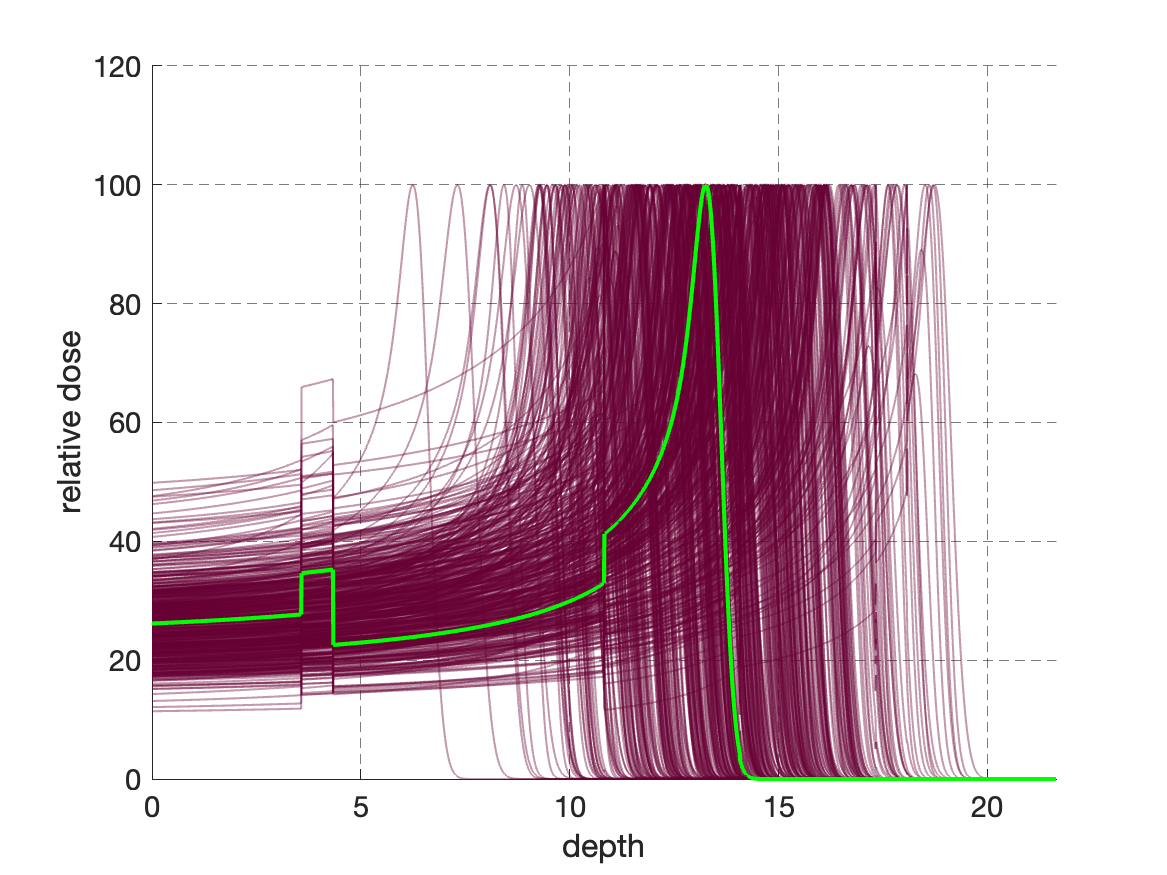}
    \caption{Posterior dose profiles ($Q$) after one SMC iteration.}
\end{subfigure}
\hfill
\begin{subfigure}{0.4\textwidth}
  \includegraphics[width=\textwidth]{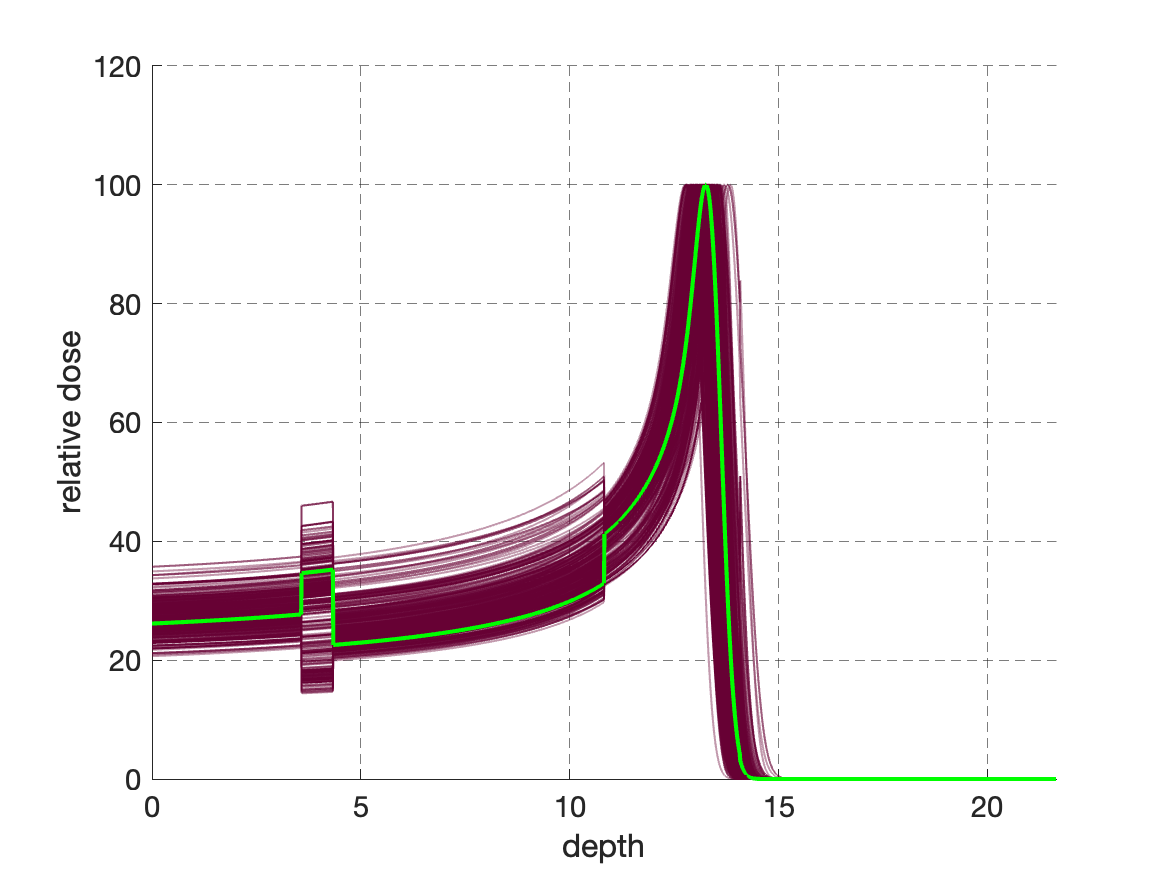}
    \caption{Posterior dose profiles ($Q$) after twenty SMC iterations.}
\end{subfigure}
\\
\begin{subfigure}{0.4\textwidth}
  \includegraphics[width=\textwidth]{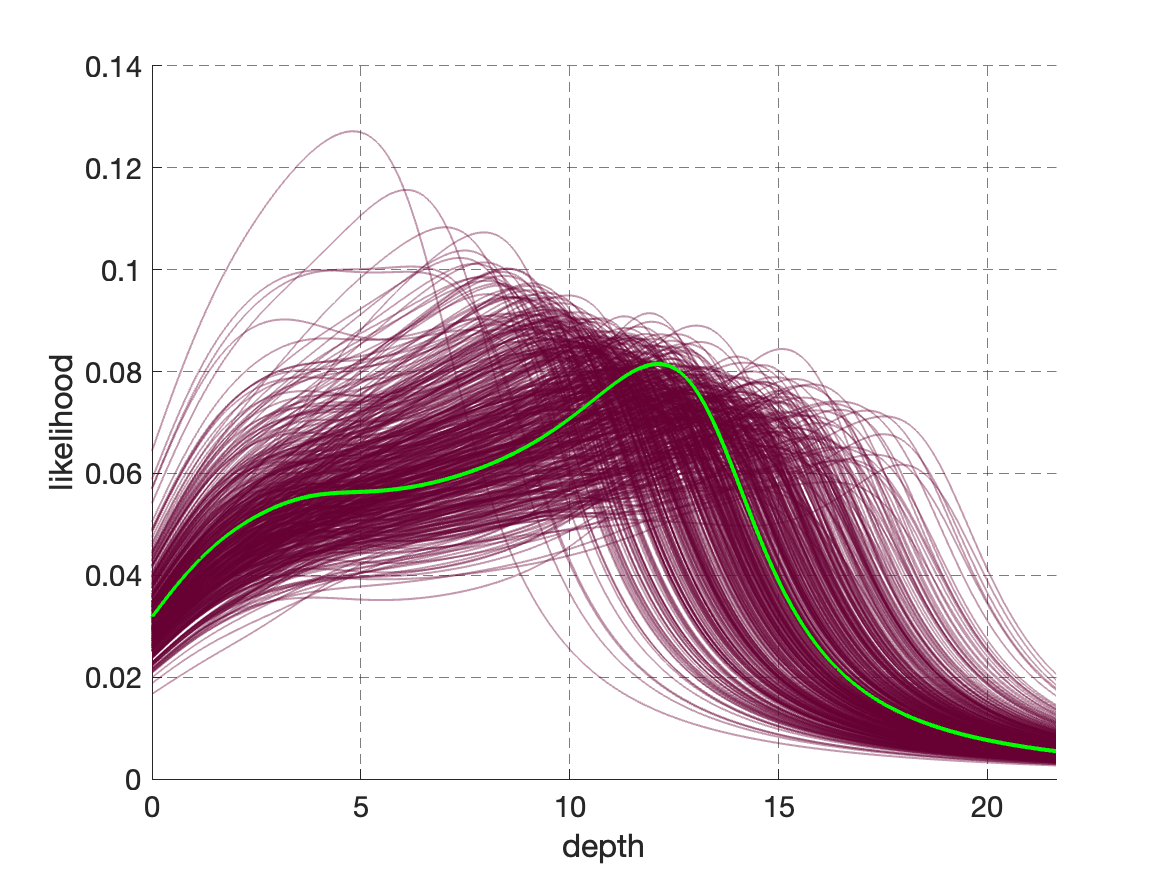}
    \caption{Posterior likelihood functions ($P$) after one SMC iteration.}
\end{subfigure}
\hfill
\begin{subfigure}{0.4\textwidth}
  \includegraphics[width=\textwidth]{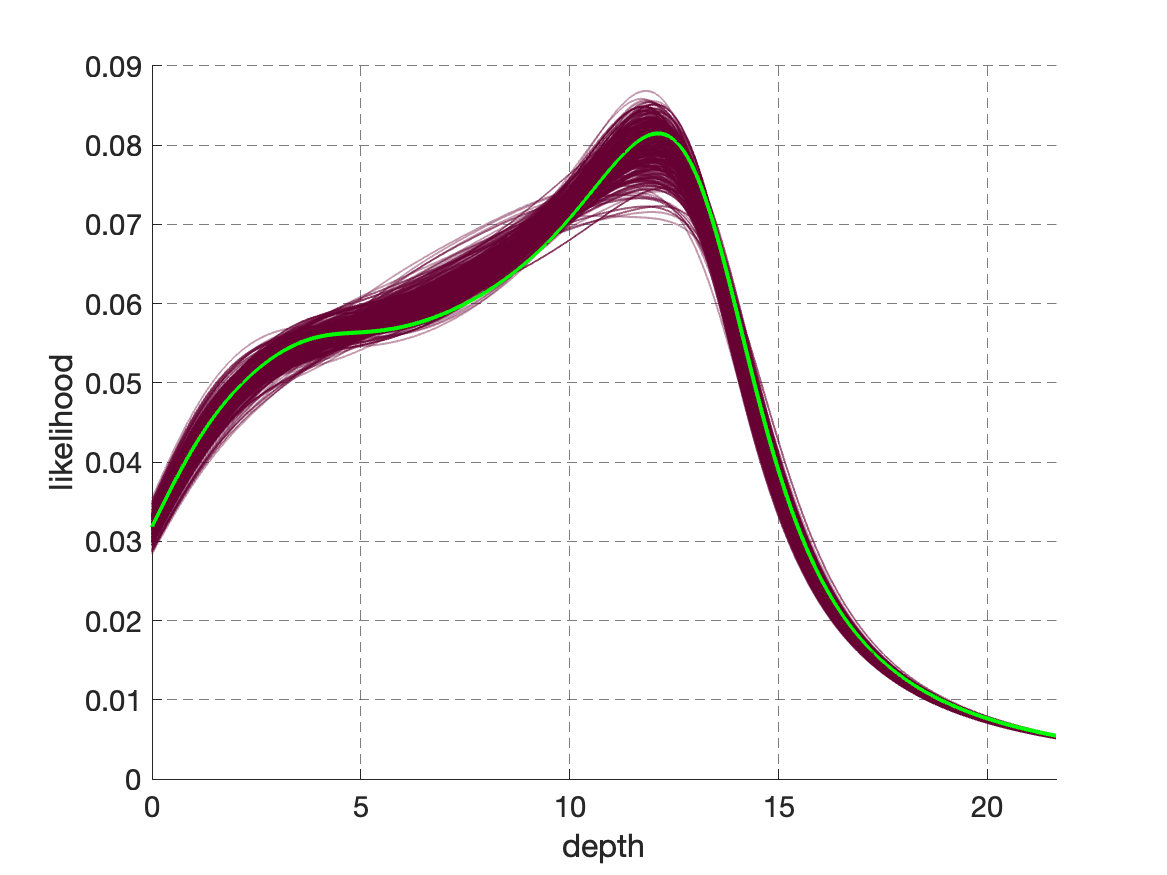}
    \caption{Posterior likelihood functions ($P$) after twenty SMC iterations.}
\end{subfigure}
\\
\begin{subfigure}{0.4\textwidth}
  \includegraphics[width=\textwidth]{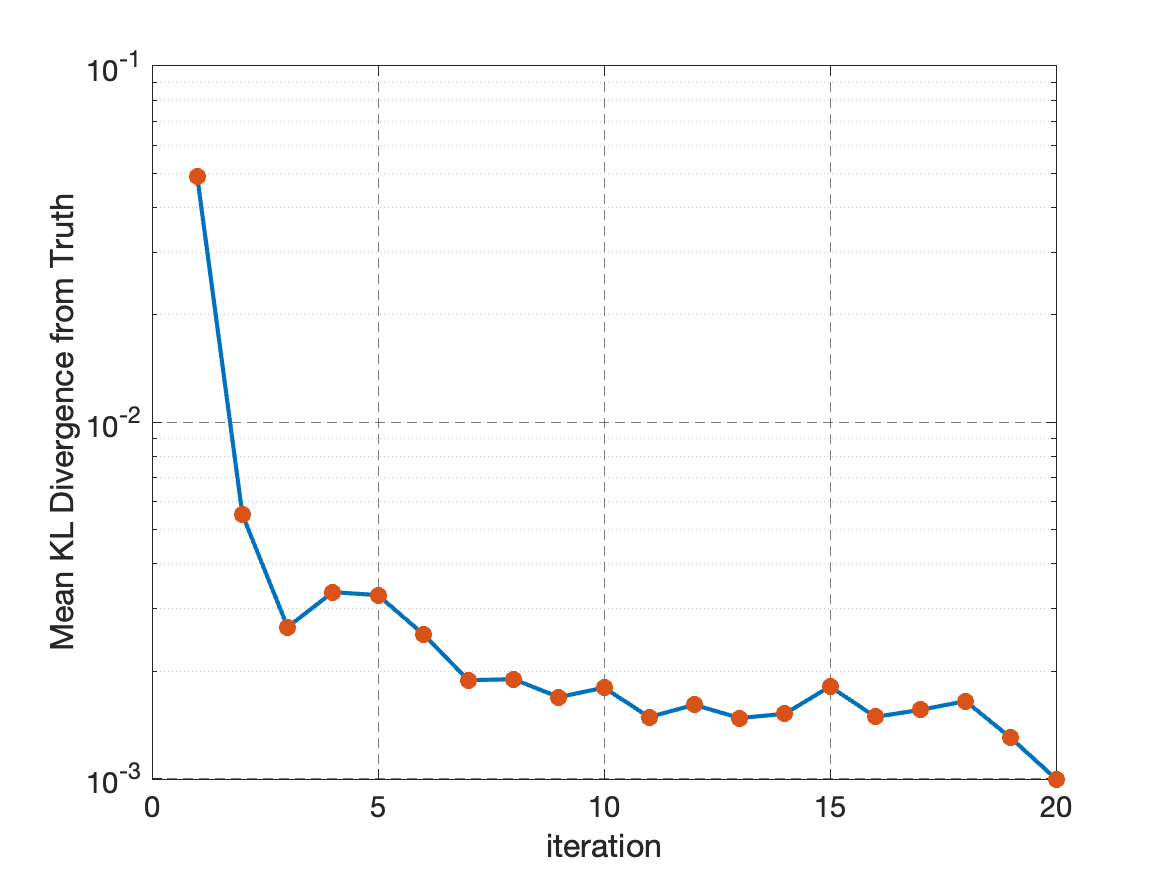}
  \caption{Kullback-Leibler divergence error as a function of SMC iterations.}
\end{subfigure}
\hfill
\begin{subfigure}{0.4\textwidth}
  \includegraphics[width=\textwidth]{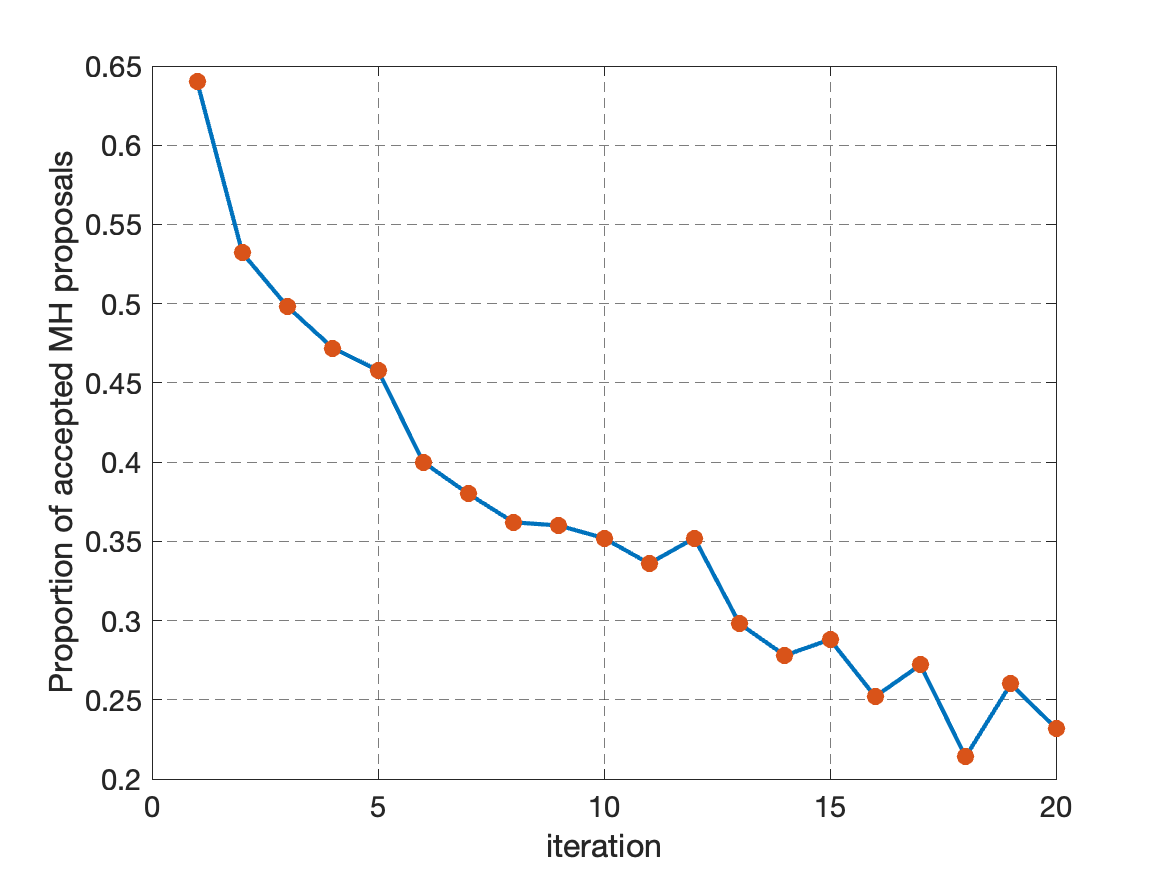}
  \caption{Proportion of accepted Metropolis Hastings proposals as a
    function of SMC iterations.}
\end{subfigure}
\caption{
  \label{fig:PBP1BLung}
  The figure shows dose profiles and likelihood
  functions associated to reconstruction of a Bragg Peak over a lung
  phantom (\S \ref{sec:lung}) from observed \pg data observed with a
  single detector. Notice the algorithm does a reasonable job in
  approximating range however struggles with differentiating between
  different tissue types.}
\end{figure*}

\begin{figure*}
\centering
\begin{subfigure}{0.4\textwidth}
  \includegraphics[width=\textwidth]{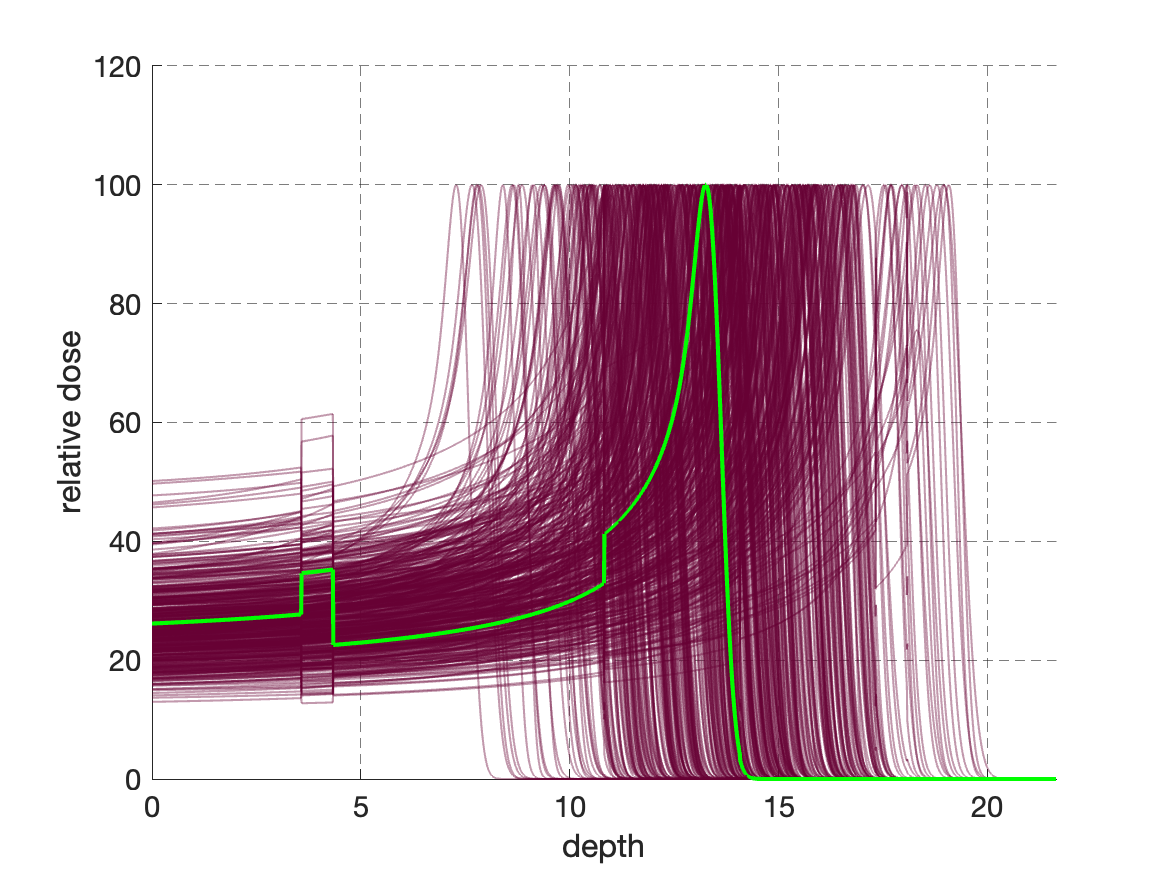}
    \caption{Posterior dose profiles ($Q$) after one SMC iteration.}
\end{subfigure}
\hfill
\begin{subfigure}{0.4\textwidth}
  \includegraphics[width=\textwidth]{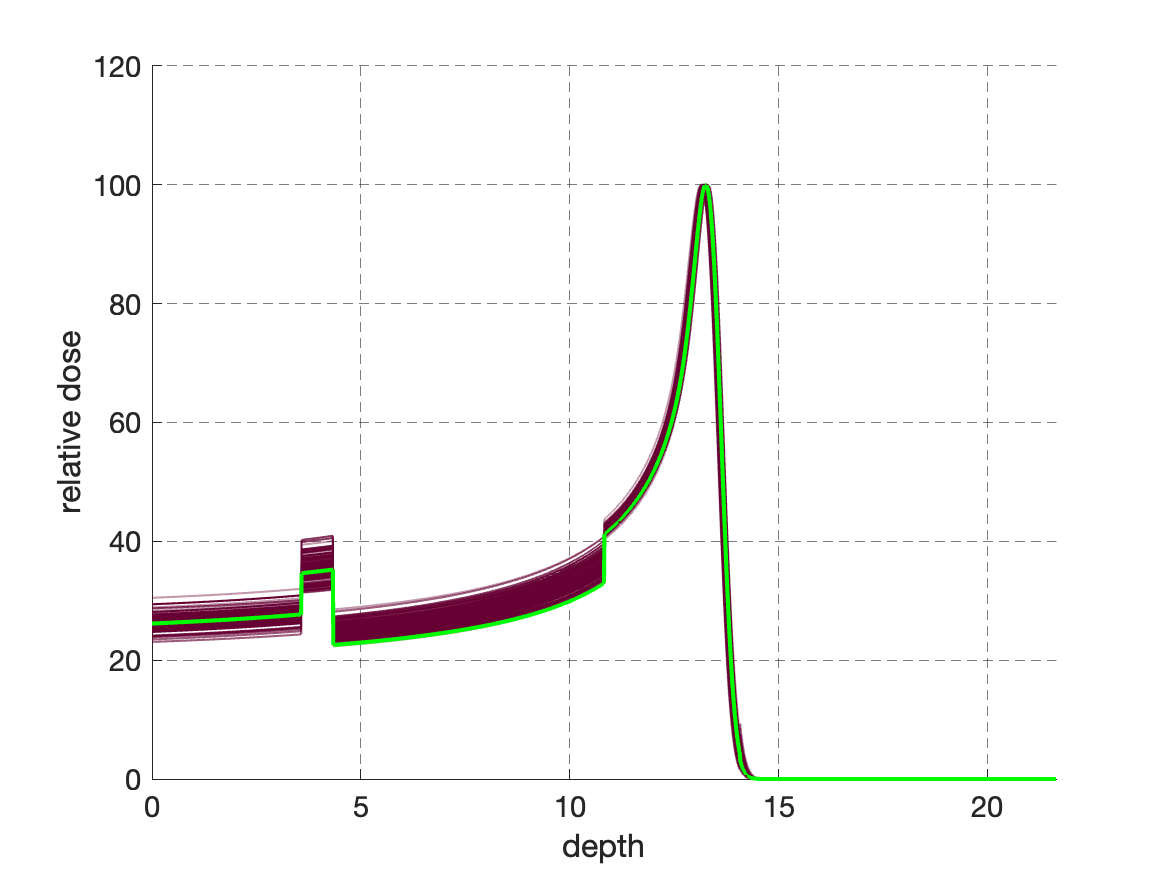}
    \caption{Posterior dose profiles ($Q$) after twenty SMC iterations.}
\end{subfigure}
\\
\begin{subfigure}{0.4\textwidth}
  \includegraphics[width=\textwidth]{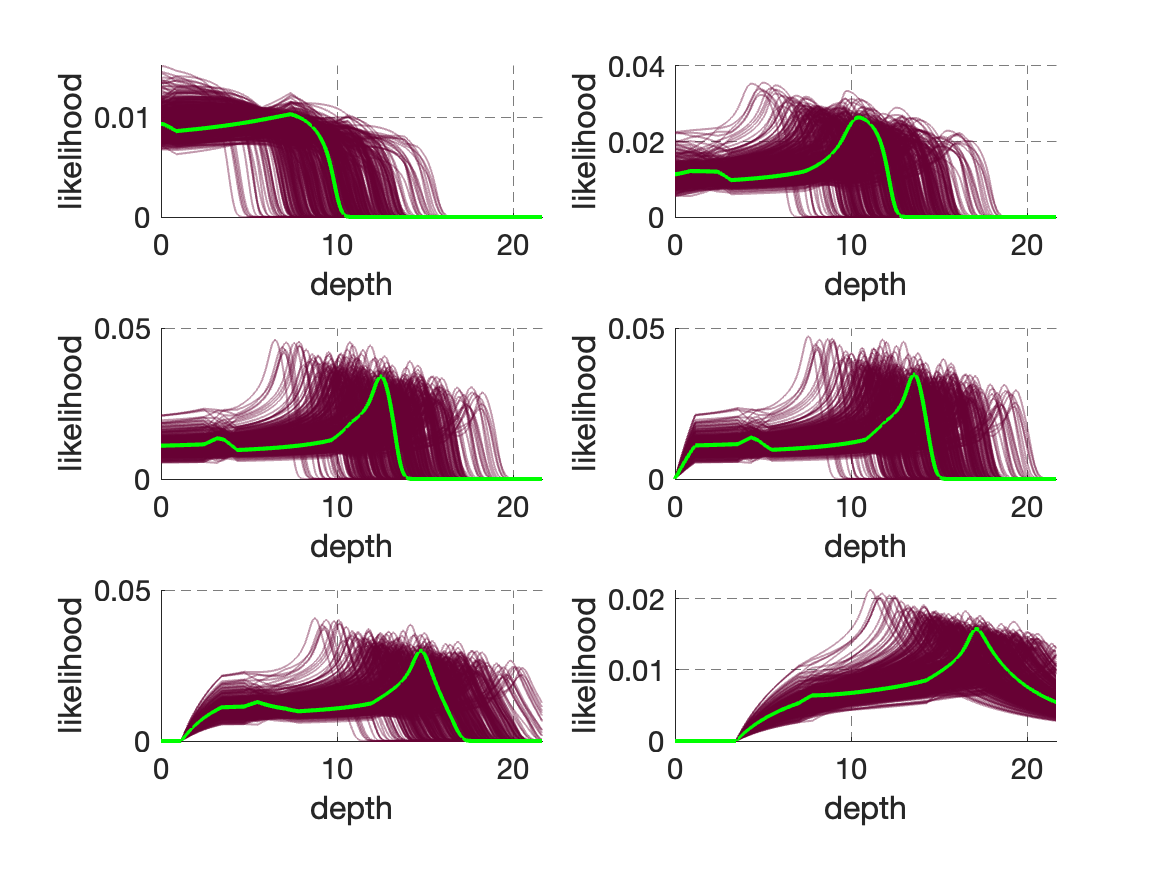}
    \caption{Posterior likelihood functions ($P$) after one SMC iteration.}
\end{subfigure}
\hfill
\begin{subfigure}{0.4\textwidth}
  \includegraphics[width=\textwidth]{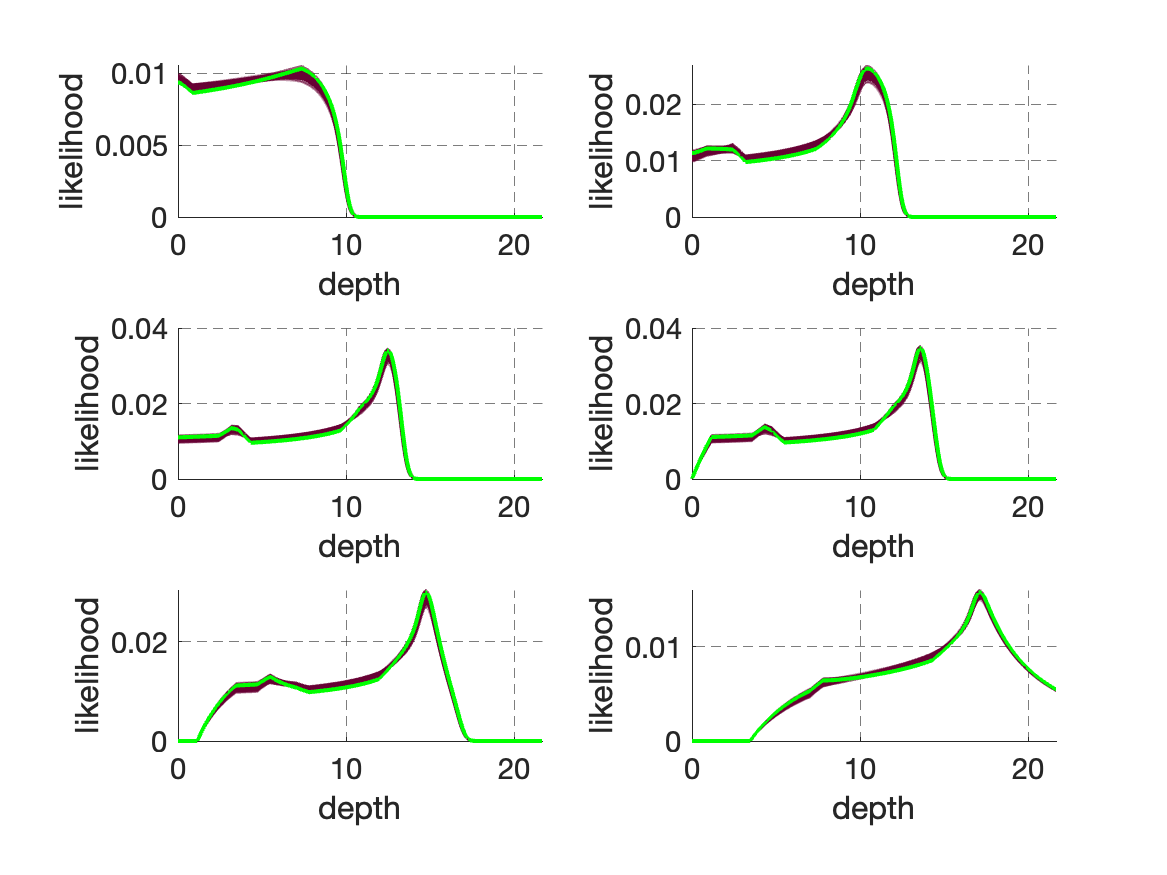}
    \caption{Posterior likelihood functions ($P$)  after twenty SMC iterations.}
\end{subfigure}
\\
\begin{subfigure}{0.4\textwidth}
  \includegraphics[width=\textwidth]{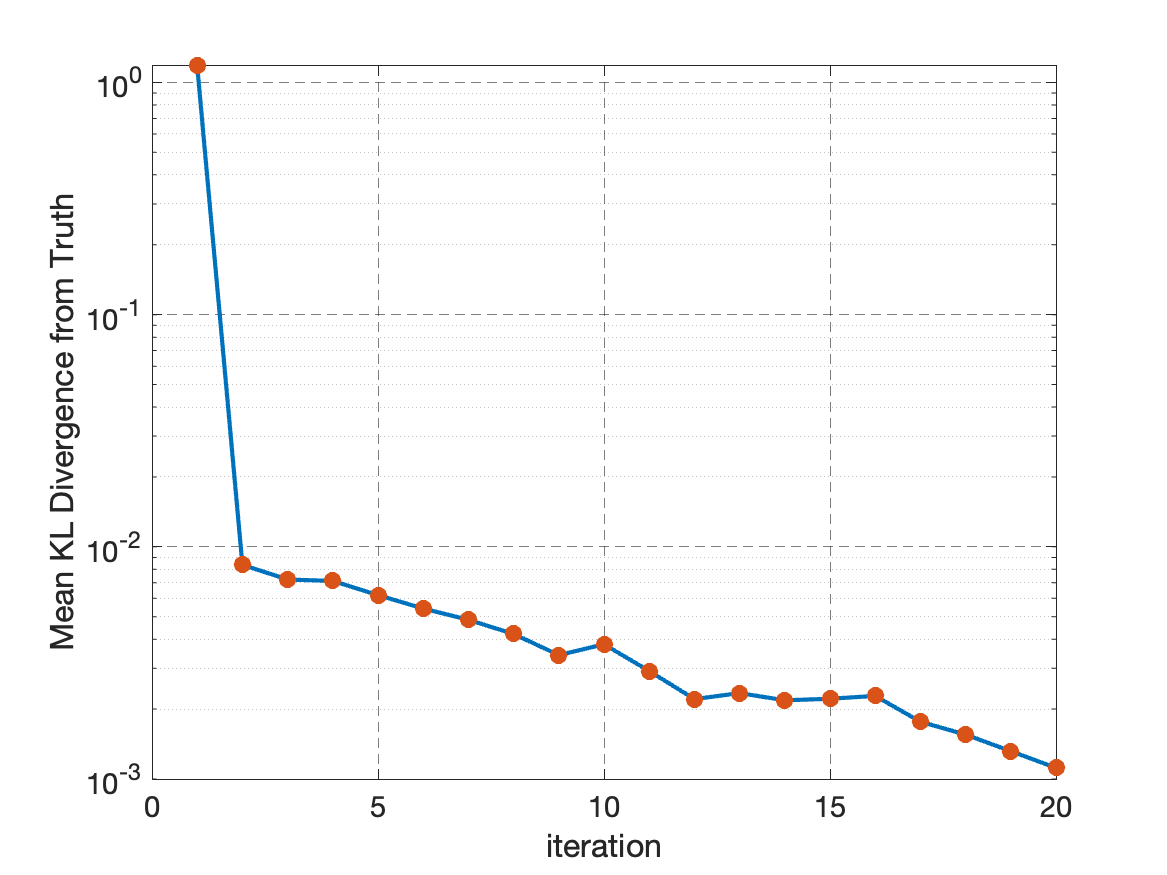}
  \caption{Kullback-Leibler divergence error as a function of SMC iterations.}
\end{subfigure}
\hfill
\begin{subfigure}{0.4\textwidth}
  \includegraphics[width=\textwidth]{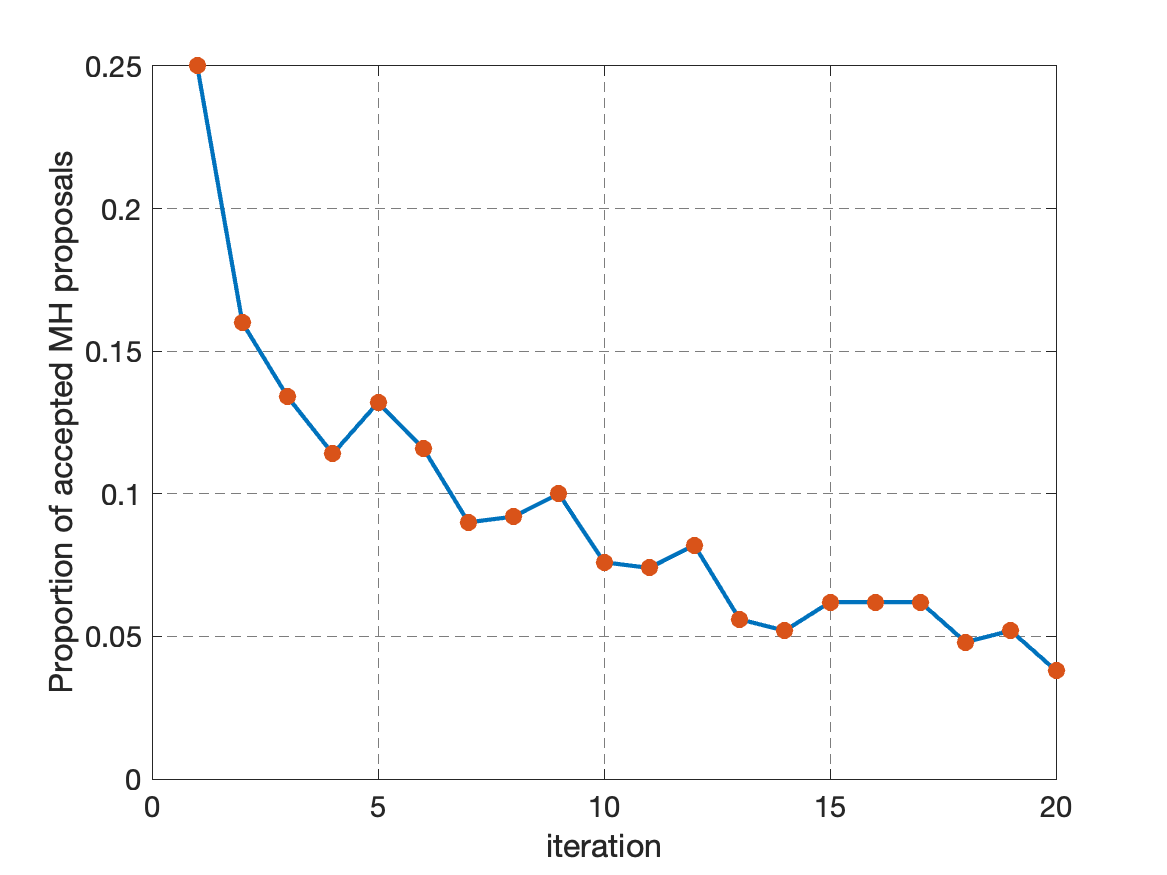}
  \caption{Proportion of accepted Metropolis Hastings proposals as a
    function of SMC iterations.}
\end{subfigure}
\caption{
  \label{fig:PBP6BLung}
  A numerical experiment showing dose profiles and likelihood
  functions associated to reconstruction of a Bragg Peak over a lung
  phantom (\S \ref{sec:lung}) from observed \pg data observed with six
  detectors. Notice the algorithm is well able to approximate the
  Bragg peak in range and is significantly more accurate in
  determining tissue profile parameters, however is not able to
  distinguish sharp transitions between tissue types of high density
  difference, in this case the transition between bone and lung.}
\end{figure*}

Our second experiment is based on a more challenging setup
representing the cross section of a lung tumour. To that end, consider
the setup of a one-dimensional cross section given in Figure
\ref{fig:lung}. Notice there are six layers between the proton source,
at a $0$cm depth and the tumorous region, between $10-13$cm although
there are eleven layers, which is important when considering the
impact of range uncertainties. We showcase the results of the
algorithm in Figures \ref{fig:PBP1BLung}--\ref{fig:PBP6BLung} where
the problem is examined with a single detector and six separate
detectors. Due to the high dimensionality of the parameter space the
additional angular resolution with six detectors is important to
determine the correct dose deposition.

\clearpage
\section{Summary and outlook}
\label{sec:summary}

In this paper we have presented a flexible mathematical framework
which could form the basis of future verification methods for PBT. In
future work, our framework will enable us to analyse the amount of
information required to measure clinically relevant differences,
potentially leading to adapted treatment plans, and offers a robust
numerical method for providing clinicians with more accurate data on
the effectiveness and precision of the delivered treatment.

We anticipate that the methods presented here will play a significant
role in the future development of $\gamma$-detectors, and we plan to
apply our methods as an integral part of this advancement in future
work. Future areas for further work include:
\begin{itemize}
\item Recent work (\cite{perez-lara_first_2023}) has developed novel
  Compton cameras for treatment verification. Compton cameras capture
  complex information about $\gamma$-radiation, which potentially
  includes some information about the direction of the observed
  photon. It should be possible to incorporate this information in a
  statistically meaningful manner due to the Bayesian elements of our
  approach. As demonstrated through consideration of different angular
  decomposition (Figure~\ref{kp}), increased angular resolution should
  enable more robust discrimination between models. Using the methods
  presented in this work should make quantifying these benefits
  straightforward.
  
\item The current approach does not consider noise in the system, due
  e.g. to background radiation. In practical situations, calibrating
  these models will be crucial for understanding different noise
  sources and incorporating them into our updating framework.
  Similarly, we assume in the setup considered in
  Figure~\ref{fig:lung} that the depths of different layers are
  known. A more realistic scenario might include uncertainty on these
  depths. The flexible nature of our approach will be important in
  developing these models.

\item An important question for design of future verification systems
  will be how to optimally place sensors. The work presented in this
  paper can address this challenge by quantifying various system
  designs.
  
\item In this paper, we use a relatively simple forward model, which
  approximates the particle flux using a one-dimensional pencil
  beam. In practice, more realistic PDE models could be implemented to
  provide more accurate results. It seems likely that these methods
  will be significantly more computationally demanding.  In the
  algorithm proposed, each new configuration requires a new numerical
  solve for the forward model at the given configuration. It seems
  likely that modified numerical methods, where new particles are
  introduced at different levels of numerical fidelity, could
  potentially enhance numerical fidelity in these cases (see, for
  example, \cite{beskos_multilevel_2017}).
  
\end{itemize}


\enlargethispage{20pt}

\paragraph*{Acknowledgements}
We would like to thank our friends and colleagues who have generously
offered their attention, thoughts and encouragement in the course of
this work. We thank Colin Baker and Sarah Osman who kickstarted this
work. All authors were supported by the EPSRC programme grant MaThRad
EP/W026899/2. Furthermore, AMGC and AEK are grateful for partial
support from EP/P009220/1. TP is grateful for partial support from
EPSRC (EP/X017206/1, EP/X030067/1) and the Leverhulme Trust
(RPG-2021-238).

\bibliographystyle{RS} 
\bibliography{ProtonTherapy} 

\end{document}